\documentclass[journal,twoside]{IEEEtran}
\usepackage{cancel}
\usepackage{tabularx,booktabs}
\usepackage{lipsum}
\newcolumntype{C}{>{\centering\arraybackslash}X} 

\usepackage{cite}

\usepackage{wrapfig}
\ifCLASSINFOpdf
   \usepackage[pdftex]{graphicx}
   \graphicspath{{../pdf/}{../jpeg/}}
   \DeclareGraphicsExtensions{.pdf,.jpeg,.png}
\else
   \usepackage[dvips]{graphicx}
   \graphicspath{{../eps/}}
   \DeclareGraphicsExtensions{.eps}
\fi
\usepackage[cmex10]{amsmath}
\usepackage{textcomp,gensymb}
\usepackage{relsize}
\usepackage{color,soul}

\usepackage{scalerel} 
\usepackage{tikz} 
\usetikzlibrary{svg.path}

\definecolor{orcidlogocol}{HTML}{A6CE39}
\tikzset{
	orcidlogo/.pic={
		\fill[orcidlogocol] svg{M256,128c0,70.7-57.3,128-128,128C57.3,256,0,198.7,0,128C0,57.3,57.3,0,128,0C198.7,0,256,57.3,256,128z};
		\fill[white] svg{M86.3,186.2H70.9V79.1h15.4v48.4V186.2z}
		svg{M108.9,79.1h41.6c39.6,0,57,28.3,57,53.6c0,27.5-21.5,53.6-56.8,53.6h-41.8V79.1z M124.3,172.4h24.5c34.9,0,42.9-26.5,42.9-39.7c0-21.5-13.7-39.7-43.7-39.7h-23.7V172.4z}
		svg{M88.7,56.8c0,5.5-4.5,10.1-10.1,10.1c-5.6,0-10.1-4.6-10.1-10.1c0-5.6,4.5-10.1,10.1-10.1C84.2,46.7,88.7,51.3,88.7,56.8z};
	}
}

\newcommand{\orcidicon}[1]{\href{https://orcid.org/#1}{\mbox{\scalerel*{
				\begin{tikzpicture}[yscale=-1,transform shape]
				\pic{orcidlogo};
				\end{tikzpicture}
			}{|}}}}

\usepackage [hidelinks] {hyperref} 

\usepackage{algorithmic}
\usepackage[ruled,vlined]{algorithm2e} 
\usepackage{amsmath,amssymb,amsfonts}
\usepackage{multicol,cuted}
\usepackage{enumitem}
\begin{document}
\title{Integration of 5G and Motion Sensors for Vehicular Positioning: A Loosely-Coupled Approach}

\author{Sharief~Saleh\textsuperscript{\orcidicon{0000-0003-1365-417X}}\,,~\IEEEmembership{Member,~IEEE,}
        Qamar~Bader\textsuperscript{\orcidicon{0000-0002-4667-1710}}\,,~\IEEEmembership{Graduate Student Member,~IEEE,}
        Malek~Karaim\textsuperscript{\orcidicon{0000-0002-2897-1330}}\,,~\IEEEmembership{Member,~IEEE,}
        Mohamed~Elhabiby\textsuperscript{\orcidicon{0000-0002-1909-7506}}\,,~\IEEEmembership{Member,~IEEE,}
        Aboelmagd~Noureldin\textsuperscript{\orcidicon{0000-0001-6614-7783}}\,,~\IEEEmembership{Senior Member,~IEEE}

\thanks{\vspace{-0pt} This work was supported by grants from the Natural Sciences and Engineering Research Council of Canada (NSERC) under grant number: ALLRP-560898-20 and RGPIN-2020-03900. (\textit{Corresponding author: Sharief~Saleh.})\\}

\thanks{Sharief Saleh was with the Navigation and Instrumentation (NavINST) Lab, Department of Electrical and Computer Engineering, Royal Military College of Canada, Kingston, ON  K7K 7B4, Canada, and is now with  Department of Electrical Engineering, Chalmers University of Technology, SE 41296 Gothenburg, Sweden (e-mail: sharief@chalmers.se).}

\thanks{Qamar Bader, Malek Karaim, and Aboelmagd Noureldin are with the Navigation and Instrumentation (NavINST) Lab, Department of Electrical and Computer Engineering, Royal Military College of Canada, Kingston, ON  K7K 7B4, Canada (e-mail: qamar.bader@queensu.ca; malekok@yahoo.com; aboelmagd.noureldin@rmc.ca).}

\thanks{Qamar Bader and Aboelmagd Noureldin are also with the Department of Electrical and Computer Engineering, Queen's University, Kingston, ON K7L 3N6, Canada.}
\thanks{Mohamed Elhabiby is with Micro Engineering Tech. Inc., and also with the Public Works Department, Ain Shams University, Cairo 11566, Egypt (e-mail: mmelhabi@ucalgary.ca).\\ \\ \\}}

\markboth{}%
{Saleh \MakeLowercase{\textit{et al.}}: Integration of 5G and Motion Sensors for Vehicular Positioning: A Loosely-Coupled Approach}



\maketitle
\bstctlcite{IEEEexample:BSTcontrol} 

\begin{abstract}

Autonomous vehicles (AVs) are poised to revolutionize the transportation industry by enhancing traffic efficiency and road safety. However, achieving optimal vehicular autonomy demands an uninterrupted and precise positioning solution, especially in deep urban environments. 5G mmWave holds immense potential to provide such a service due to its accurate range and angle measurements. Yet, as mmWave signals are prone to signal blockage, severe positioning errors will occur. Most of the 5G positioning literature relies on constant motion models to bridge such 5G outages, which do not capture the true dynamics of the vehicle. Few proposed methodologies rely on inertial measurement units (IMUs) to bridge such gaps, where they predominantly use tightly coupled (TC) integration schemes, introducing a nonlinear 5G measurement model. Such approaches, which rely on Kalman filtering, necessitate the linearization of the measurement model, leading to pronounced positioning errors. In this paper, however, we propose a loosely coupled (LC) sensor fusion scheme to integrate 5G, IMUs, and odometers to mitigate linearization errors. Additionally, we propose a novel method to design the process covariance matrix of the extended Kalman filter (EKF). Moreover, we propose enhancements to the mechanization of the IMU data to enhance the standalone IMU solution. The proposed methodologies were tested using a novel setup comprising 5G measurements from Siradel's S\_5G simulation tool and real IMU and odometer measurements from an hour-long trajectory. The proposed method resulted in 14 cm of error for 95\% of the time compared to 1 m provided by the traditional constant velocity model approach.
\end{abstract}

\begin{IEEEkeywords}
5G; autonomous vehicles (AVs); INS; Kalman filter (KF); loosely coupled (LC); mmWave; positioning.\vspace{10pt}
\end{IEEEkeywords}


\section{Introduction}
\IEEEPARstart{A}{chieving} the highest levels of autonomy within autonomous vehicles (AVs) is a crucial step towards promoting a safer driving environment. Moreover, it will increase the efficiency of the transportation network of cities, enhance their reliability, and decrease congestion rates significantly \cite{AVCongestion}.  Intelligent Transportation Systems (ITS) play a vital role in enabling this advancement by integrating various technologies and solutions to improve traffic management, enhance safety, and optimize mobility \cite{guerrero-ibanez_sensor_2018}. The Society of Automotive Engineers (SAE) had established six levels of vehicular autonomy. Level zero corresponds to having no driving automation, i.e., full human control, while level five corresponds to full automation with no need for human intervention \cite{SAE}. To achieve the highest levels of autonomy, an uninterrupted positioning solution that can reliably achieve a decimeter level of accuracy in deep-urban environments is required \cite{AVRequirements}. 
\IEEEpubidadjcol  
\noindent Nowadays, vehicles are equipped with a plethora of positioning-capable technologies such as global navigation satellite system (GNSS) receivers, network-based receivers, onboard inertial navigation systems (INS), and odometers. In some modern vehicles, a suite of perception-based systems such as cameras, lidars, and/or radars are also present. Hence, over the past decades, researchers have been researching these various positioning technologies and fusing multiple of them together to meet the AV's market positioning demands \cite{yeong_sensor_2021}. As of today, no positioning technology can offer an accurate and precise solution for all possible driving dynamics in all environments (e.g. open sky, urban canyons, covered parking garages, etc) \cite{yeong_sensor_2021}. For instance, GNSS may offer a reliable centimeter-level of accuracy in open sky environments, yet, its performance deteriorates significantly when the user equipment (UE) enters an urban canyon, where signal multipath and blockage are highly probable \cite{ProfBook}. On the other hand, dead-reckoning-based systems like inertial measurement units (IMU) and odometers would not suffer in such scenarios, as they are self-contained and environment-independent. Yet, they are constantly plagued with an accumulation of errors due to their inherent design \cite{ProfBook}. Perception-based systems may work in downtown scenarios where GNSS fails, but their performance may degrade when the scene is partially overshadowed or when highly light-reflective materials are present in the case of cameras \cite{Camera}. Also, their performance may deteriorate significantly due to weather conditions such as rain and snow \cite{Radar}. Other perception systems that are weather-independent, like radars, may suffer from a short range of operation and relatively low resolution \cite{Radar}.

The emerging fifth-generation (5G) New Radio (NR) cellular network, on the other hand, is equipped with a plethora of key technologies that can support high-precision positioning in urban canyons. Moreover, unlike perception-based systems, 5G NR is not affected by lightning conditions. Additionally, 5G referencing signals have a bandwidth of $400$ MHz, resulting in notably accurate time-based measurements \cite{wymeersch_integration_2021}. Furthermore, 5G systems are equipped with massive MIMO capabilities, enabling precise downlink angle of departure (DL-AOD) and uplink angle of arrival (UL-AOA) measurements \cite{merits}. Thus, 5G NR is anticipated to take the role of GNSS satellites in urban canyon environments, by providing position-fixing services. 

Like real-life GNSS implementations, 5G positioning services are anticipated to be fused with other technologies. Such fusion can be done in a loosely coupled (LC) or a tightly coupled (TC) fashion. LC fusion schemes, also known as decentralized integration, utilize the end result of the fused systems to find a better solution \cite{ProfBook}. For instance, an LC integration between INS and 5G would fuse the mechanized position, velocity, and attitude outputs of the INS and the positioning solution of 5G. The LC integration scheme is desirable for its simple implementation and robustness \cite{ProfBook}. On the other hand, TC fusion schemes, also known as centralized integration, utilize a single/central integration filter to fuse the raw measurements of the integrated technologies \cite{ProfBook}. Therefore, the integrated technologies do not need to estimate the positioning states on their own. This is advantageous for trilateration/triangulation-based technologies (like GNSS), as they no longer require 3-4 wireless nodes to position the vehicle \cite{ProfBook}. However, TC implementations are far more complex than their LC counterparts \cite{ProfBook}. Moreover, the utilization of raw 5G measurements will lead to a non-linear measurement model, which is problematic for linear filters like the Kalman filter. Unlike the GNSS-INS research literature, the integration of the 5G positioning technology with other onboard motion sensors (OBMS) hasn't been sufficiently investigated. This work aims to address this gap to provide an uninterrupted high-precision positioning solution via the integration of 5G positioning technology with low-cost OBMS. Toward that end, the paper offers the following contributions:
\begin{itemize} [leftmargin=*]
    \item An automatic mechanization-stopping mechanism to reduce the accumulation of OBMS errors.
    \item A novel approach for designing the covariance and noise coupling matrices for the extended Kalman filter (EKF).
    \item An LC integration scheme between the 5G positioning solution and OBMS using an EKF.
    \item A novel experimental setup for conducting tests using quasi-real 5G and real OBMS measurements.
\end{itemize}


\section{Literature Survey}\label{LR}
Most of the 5G positioning literature is focused on the snapshot positioning problem, i.e., estimating the static state of the UE, with minimal attention given to the tracking problem, i.e., estimating the dynamic state of the UE \cite{mogyorosi_positioning_2022}. In reality, the UE is expected to be in motion, and thus, the tracking problem becomes more relevant, compared to snapshot positioning. Usually, the UE has access to a motion model, which will greatly improve the accuracy of the positioning estimate. In the 5G tracking literature, the usage of constant motion-based transition models like constant velocity/acceleration models has been prevalent, \cite{Fokin2021,saleh_vehicular_2021,wen2024high,koivisto_high-efficiency_2017}, as opposed to measurement-based transition models that rely on external sensors like IMUs \cite{mostafavi_vehicular_2020, luo_research_2021,wang_simulation_2022}. As the name suggests, constant motion models are best suited for applications where the velocity/acceleration of the UE is expected to be constant from one epoch to the next. Yet, as UEs mounted on AVs experience dynamic changes in acceleration, speed, and orientation, the constant motion models will lead to high positioning errors due to model mismatch. On the other hand, measurement-based transition models rely on the sensed acceleration and angular velocity of the UE to predict its future state, which is more accurate. All of the aforementioned works that utilize measurement-based transition models use TC integration schemes. In the following, we will briefly present their contributions. Authors in \cite{mostafavi_vehicular_2020} have proposed a tightly coupled integration between 5G's AOD and TOA measurables and two accelerometers mounted on the x and y axes of the vehicle. The proposed method utilizes a constant acceleration model and utilizes the accelerometer measurements in the correction stage of the EKF. The methodology was tested using simulated data for both the 5G and the IMU. The integration with accelerometers alone without adding gyroscope measurements is not traditional, as such sensors are usually packed together. Moreover, the simulation was done in 2D, hence, neglecting the estimation of the pitch and roll angles. Additionally, the states of the proposed filter also neglect the estimation of the azimuthal angle of the vehicle and the biases of the IMUs, which is crucial for real-life AV operation. Finally, the use of a TC scheme will incur high linearization errors in the range and angle measurements from the 5G side. The work reported in \cite{luo_research_2021} proposes an invariant extended Kalman filter (InEKF) approach to the integration problem. The proposed method utilizes simulated vertical and horizontal AODs and TOA measurements. Authors in this work assume synchronization between base stations (BSs) and the UE, which is typically not guaranteed. The proposed method works in 3D and utilizes six simulated IMU sensors to conduct the state transition process, which is advantageous compared to the previous work's constant acceleration model. Moreover, the states of the filter encompass position, velocity, and attitude (PVA) navigation states as well as the six IMU biases, which is traditionally done in GNSS-INS filters. The method proposes a 5G hybrid positioning scheme that utilizes a centralized TC-based least-squares (LS) approach to integrate between the available BSs which results in linearization errors. Authors in \cite{wang_simulation_2022} integrate simulated measurements from 5G, INS, GNSS, and low-earth orbit (LEO) satellites using three TC sub-filters and a centralized fusion filter. The proposed 5G-INS sub-filter utilizes TOA measurements from four BSs to conduct centralized trilateration, which imposes two main challenges. First, having access to four BSs is challenging in real-life scenarios. Second, trilateration will cause high linearization errors as the UE is in close vicinity to one of the connected BSs \cite{saleh_vehicular_2021,saleh_5g-enabled_2022}. The method also estimates time bias between the UE and the BS, which makes the use of TOA acceptable. The proposed method utilizes a linearized state transition model which will cause unnecessary linearization errors. In conclusion, all of the proposed methods utilize TC schemes that incur unnecessary linearization errors and utilize simulated 5G and IMU data, which does not capture the true challenges faced in real-world scenarios.

\section{System Model}
\subsection{Positioning Problem Formulation}
In this work, the 3D position of the UE is denoted as $(x,y,z)$ in the universal transverse Mercator (UTM) coordinate system and as $(\varphi,\lambda,h)$ in the world geodetic system (WGS). Here, $(\varphi,\lambda,h)$ are the latitude, longitude, and altitude of the vehicle, respectively. The position of the BSs, on the other hand, are denoted as $(x_{BS},y_{BS},z_{BS})$. The euclidean distance between the UE and the BS can thus be formulated as
\begin{equation}\label{Range}
    r=\sqrt{\Delta x^2+ \Delta y^2 + \Delta z^2},
\end{equation}
where $(\Delta x, \Delta y, \Delta z)$ denote the difference between the UE's position $(x,y,z)$ and the wireless node's position $(x_{BS},y_{BS},z_{BS})$. The relative horizontal and vertical angles between the BS and the UE in the UTM reference frame are formulated as
\begin{equation}\label{Angle}
    \begin{split}
        \theta&=\tan^{-1}\left(\frac{\Delta y}{\Delta x}\right),\\
        \psi&=\sin^{-1}\left(\frac{\Delta z}{r}\right),
    \end{split}
\end{equation}
where $\theta$ is the counterclockwise horizontal angle with respect to the x-axis and $\psi$, is the counterclockwise vertical angle with respect to the xy-plane. The velocity states of the vehicle represented in the local-level frame (l-frame) are expressed as $(v_e, v_n, v_u)$, which denotes the velocities in the east, north, and up directions, respectively. The vehicle velocities denoted in the body frame (b-frame) of the vehicle are expressed as $(v_x,v_y,v_z)$, which correspond to velocities in the lateral, longitudinal, and vertical directions, respectively. The angular rotations between the vehicle's b-frame and the l-frame constitute the attitude state of the vehicle. The attitude angles are the pitch, roll, and azimuth angles, which are expressed as $(p,r,a)$, respectively.

\subsection{5G Measurables} 
5G NR features uplink sounding reference signal (UL-SRS) and downlink positioning reference signal (DL-PRS) dedicated for UE positioning \cite{refsignals}. Both signals are utilized to compute time-based range measurements and angle-based measurements. In this work, round-trip-time (RTT) and DL-AOD measurables are utilized.
%
%
We also consider that BSs have access to a uniform linear array (ULA) of antennas, enabling horizontal AOD measurements only. To compute the 3D position of the UE, an assumption should be made about the height of the UE. That is, we assume that the UE maintains a constant height, which is a fair assumption for AVs. Therefore, the position of the UE can be computed as
\begin{equation}\label{2D Pos}
    \begin{split}
        &x=r_{2D} \cdot \sin(\theta) +x_{BS},\\
        &y=r_{2D} \cdot \cos(\theta) +y_{BS},\\
        &r_{2D}=\sqrt{r^2-\Delta z^2},
    \end{split}
\end{equation}
where $r_{2D}$ is the 2D range between the BS and the UE.

\subsection{INS Measurables}
Dead-reckoning systems rely on initial PVA states and accumulate measurements from inertial sensors, e.g., accelerometers, gyroscopes, and odometers, to estimate the vehicle's current PVA states \cite{ProfBook}.

\subsubsection{Accelerometers and Gyroscopes}
Three orthogonally mounted accelerometers, $f_b=[f_x, f_y, f_z]^{\top}$, and gyroscopes, $\omega_b~=~[\omega_x, \omega_y, \omega_z]^{\top}$, are used to measure the acceleration and rate of rotation of the vehicle around its axes. These acceleration and angular velocity measurements are in the vehicle's body frame in order to utilize them for navigation purposes in a global reference frame, thus, they need to be transformed into the l-frame. Such transformation is conducted via the $\boldsymbol{R}_b^l$ rotation matrix defined in \cite{ProfBook}. 
%
%
The transformed l-frame accelerometer and gyroscope measurements are defined as
\begin{equation}\label{LocalMeasurements}
    \begin{split}
        f_l&=\boldsymbol{R}_b^l f_b,\\
        \omega_l&=\boldsymbol{R}_b^l \omega_b,\\
    \end{split}
\end{equation}
where $f_l=[f_e, f_n, f_u]^{\top}$ are the east, north, and up accelerations, respectively; and  $\omega_l=[\omega_p, \omega_r, \omega_a]^{\top}$ are the rates of rotation of the pitch, roll, and azimuth angles, respectively.

\subsubsection{Odometers}
Wheel odometer sensors are used with wheel-based vehicles to compute the forward velocity of the vehicle, in the b-frame \cite{ProfBook}. This is done by measuring the number of turns the wheel turned per unit time; $\omega_{\text{Odo}}$. The resulting forward velocity, $v_y=v_{\text{Odo}}$, is computed as follows 
\begin{equation}\label{odo}
        v_{\text{Odo}}= 2\pi r_{\text{wheel}} \cdot \omega_{\text{Odo}},
\end{equation}
where $r_{\text{wheel}}$ is the radius of the wheel. Like the inertial sensors, the rotation matrix is used to compute the l-frame velocities as follows:
\begin{equation} \label{Vel}
\begin{bmatrix}
v_e\\
v_n\\
v_u
\end{bmatrix} = 
\begin{bmatrix}
\sin{a}\cos{p}\\
\cos{a}\cos{p}\\
\sin{p}
\end{bmatrix} v_{\text{Odo}}.
\end{equation}

\section{Proposed LC Integration of 5G mmWave with Onboard Motion Sensors}
\begin{figure}[t!]
	\centering
	\includegraphics[width=\columnwidth,trim=15 18 15 15,clip]{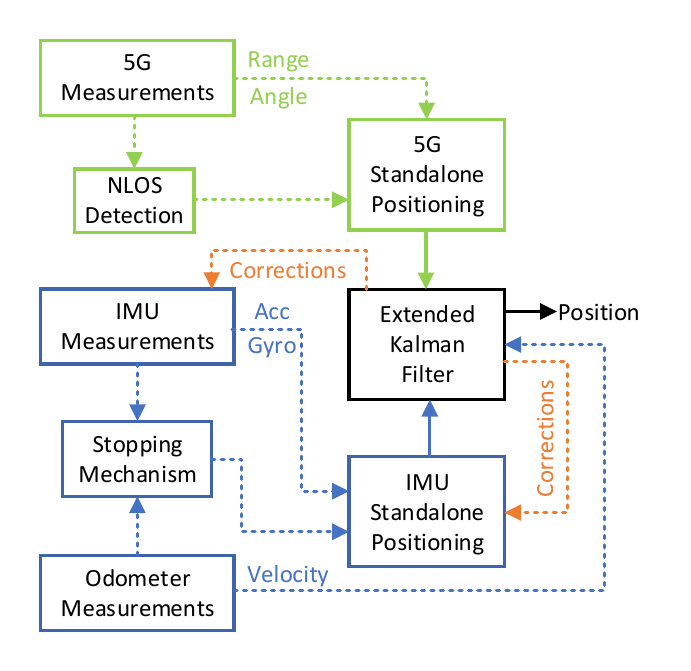}
	\DeclareGraphicsExtensions.
	\caption{Proposed LC integration between 5G and OBMS.}
	\label{Proposed Methods}
\end{figure}

In this section, we propose an LC integration between 5G, a six-degrees-of-freedom (DoF) IMU, and an odometer as shown in Fig. \ref{Proposed Methods}. Here, we utilize a non-line-of-sight (NLOS) detection methodology, proposed in \cite{bader_nlos_2022}, to exclude NLOS BSs and to determine when to fully switch to standalone INS operation. Moreover, we argue that the use of LC integration will mitigate the unnecessary linearization errors encountered by TC schemes found in the literature. Additionally, we introduce a new framework for the process covariance matrix $\boldsymbol{Q}$ and the noise coupling matrix $\boldsymbol{G}$ to reduce the filter tuning burden. We also present the nuances of the proposed mechanization scheme which includes state initialization, automatic bias removal, and automatic halting mechanisms to reduce IMU bias accumulation while maintaining an agile dynamic response.

\subsection{INS Preprocessing} 
To compute the 9D PVA states using INS, multiple stages of: state initialization, IMU bias estimation, INS mechanization, and error correction, should take place.

\subsubsection{INS State Initialization}
INS navigation is a dead-reckoning positioning scheme, where inertial measurements are accumulated on top of already established navigation states. Accurate initialization of these nine navigation states $[\varphi,\lambda,h,v_e,v_n,v_u,p,r,a]^{\top}$ is paramount, as the slightest error in them will accumulate over time, leading to positioning errors. In this paper, the position states are initialized with the aid of 5G measurements, and the initial velocity states are set to zero, as the vehicle is expected to be at a resting state at the start of the trajectory. Finally, the orientation of the vehicle is initialized via the reference solution, as orientation estimation through 5G requires a 2D array at the UE \cite{shahmansoori_position_2018}, which is not assumed in this paper.

\subsubsection{Initial IMU Bias Estimation}
At the start of a trajectory, initial biases can be estimated with relative ease given that the vehicle is stationary at the time it was turned on for a known amount of time $T$. To remove gyroscopes' initial biases, the means of the gyroscopes' measurements over the period $T$ are computed and assumed to represent the initial biases for the respective gyroscopes as seen in (\ref{Initial Gyro Biases}). This method can be employed repeatedly within the trajectory whenever the vehicle is deemed to be not in motion. A method of vehicle motion detection will be discussed in Section \ref{Stopping Mech}.
\begin{equation}\label{Initial Gyro Biases}
    \begin{bmatrix}
        \delta \omega_x\\
        \delta \omega_y\\
        \delta \omega_z\\
    \end{bmatrix}=\frac{1}{T} \sum\limits^T
    \begin{bmatrix}
        \boldsymbol{\omega}_x\\
        \boldsymbol{\omega}_y\\
        \boldsymbol{\omega}_z\\
    \end{bmatrix}
\end{equation}
On the other hand, accelerometer biases are not as easy to estimate, as the mean values of their measurements over the period $T$ do not only consist of biases but also include a gravitational force component. The extent to which the gravitational component is present at the individual accelerometers depends on the orientation of the vehicle. Therefore, the initial attitude of the vehicle is utilized to project the l-frame gravitational acceleration $[0,0,g]^{\top}$ to b-frame accelerations $[g_x,g_y,g_z]^{\top}$ via the $\boldsymbol{R}_b^l$ matrix \cite{ProfBook}. The difference between the mean of the accelerometer measurements over the period $T$ and the projected gravitational accelerations are then assumed to represent the biases in the respective accelerometers as seen in (\ref{Initial Acc Biases}). It is worth noting that attempting to remove accelerometer biases whenever the vehicle is at rest will highly depend on the quality of the orientation estimates. Thus, it is not advised to manually reset such errors, unlike the case with gyroscopes.


\begin{equation}\label{Initial Acc Biases}
    \begin{bmatrix}
        \delta f_x\\
        \delta f_y\\
        \delta f_z\\
    \end{bmatrix}=\frac{1}{T} \sum\limits^T
    \begin{bmatrix}
        \boldsymbol{f}_x\\
        \boldsymbol{f}_y\\
        \boldsymbol{f}_z\\
    \end{bmatrix} - \boldsymbol{R}_l^b \begin{bmatrix}
         0\\ 0\\ g
    \end{bmatrix}
\end{equation}

\subsection{INS Mechanization}
\begin{figure}[t!]
	\centering
	\includegraphics[width=\columnwidth]{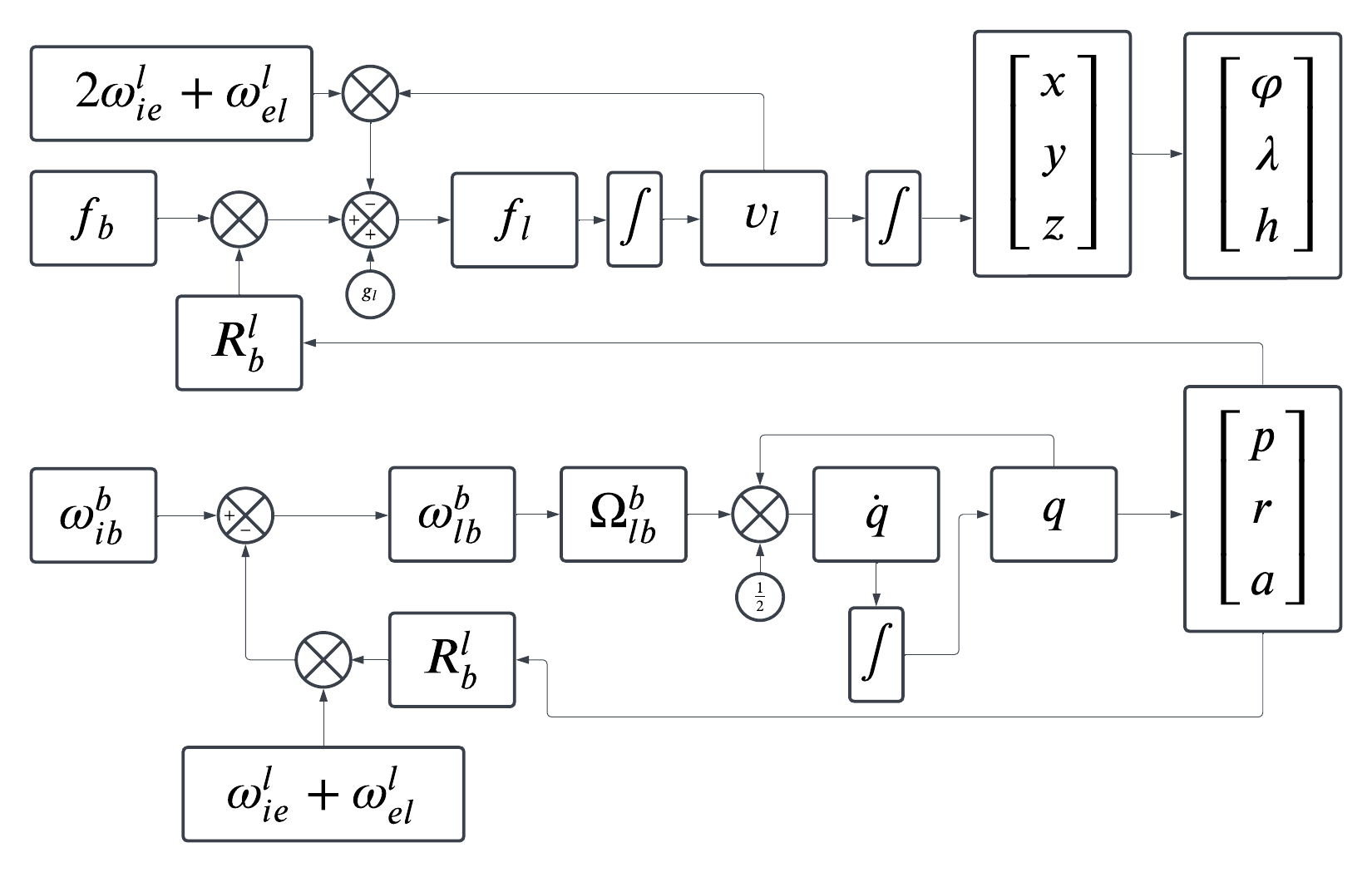}
	\DeclareGraphicsExtensions.
	\caption{INS mechanization block diagram.}
	\label{INS Mech}
\end{figure}

INS mechanization is the process of computing the PVA states by accumulating and transforming acceleration and gyroscope measurements. The general INS mechanization block diagram, shown in Fig. \ref{INS Mech}, constitutes two main steps. First, computing the attitude angles by removing extraneous effects and accumulation of gyroscope measurements. Second, computing velocities and position in the l-frame by measurement projection, removing extraneous effects, and accumulation of the resultant measurements (twice for a position estimate).

\subsubsection{Computing Attitude}
Gyroscope measurements consist of three main components; aside from biases and noise. Namely, the Earth's angular velocity with respect to the inertial frame ($15$ deg/hr), the orientation change due to motion on curvilinear coordinates in the local-level frame, and the true angular velocity of the vehicle's body with respect to  the local-level frame as seen in (\ref{Gyro Components}). In order to compute the attitude of the vehicle, the first two components should be estimated and removed. 

\begin{equation}\label{Gyro Components}
        \boldsymbol{\omega}_{lb}^b = \boldsymbol{\omega}_{ib}^b-\boldsymbol{R}_l^b (\boldsymbol{\omega}_{ie}^l + \boldsymbol{\omega}_{el}^l)
\end{equation}

\noindent The earth rotation component, $\boldsymbol{\omega}_{ie}^l=\boldsymbol{R}_e^l\begin{bmatrix}0 & 0 & \omega^e \end{bmatrix}^{\top}$, comprises the angular velocity between the Earth-centered Earth-fixed frame (e-frame) and the i-frame along the north-pole \cite{ProfBook}. The rotation matrix from the e-frame to the l-frame is explained in \cite{ProfBook}.
\noindent The second component, $\boldsymbol{\omega}_{el}^l$, is computed via the knowledge of velocities of the vehicle in the east, north, and up directions:  \cite{ProfBook}
\begin{equation}\label{wel}
\boldsymbol{\omega}_{el}^l = \begin{bmatrix}
-\frac{v_n}{M + h} &
\frac{v_e}{N + h} &
\frac{v_e \tan(\varphi)}{N + h}
\end{bmatrix}^{\top},
\end{equation}
where $M$ and $N$ are Earth's  meridional and  prime-vertical radii, respectively, and are described in \cite{ProfBook}. The attitude of a vehicle can be computed from $\boldsymbol{\omega}_{lb}^b$ using quaternions. The initial attitude of the vehicle can be represented in a quaternion form, denoted by $\boldsymbol{q}_0$. A quaternion vector can be expressed in terms of the $\boldsymbol{R}_b^l$ matrix and vice versa as explained in \cite{ProfBook}. At each time step, the quaternion rate of change due to $\boldsymbol{\omega}_{lb}^b$, denoted by $\boldsymbol{\dot{q}}_k$, is calculated as follows: \cite{ProfBook}
\begin{equation}\label{qdot}
    \boldsymbol{\dot{q}}_k = \frac{1}{2} \boldsymbol{\Omega}_{lb_k}^b \boldsymbol{q}_k,
\end{equation}
where $\boldsymbol{\Omega}_{lb_k}^b$ is a skew matrix constructed by the vector $\boldsymbol{\omega}_{lb}^b$. The updated quaternion, denoted by $\boldsymbol{q}_{k+1}$, is obtained by integrating the quaternion rate of change as shown in (\ref{Quaternion}) \cite{ProfBook}.
\begin{equation}\label{Quaternion}
    \boldsymbol{q}_{k+1} = \boldsymbol{q}_k +  \boldsymbol{\dot{q}}_k\Delta t
\end{equation}

\subsubsection{Computing Velocities and Position}
Like gyroscope measurements, the accelerometers' extraneous components comprise the earth's rotation component and an orientation change due to velocity on curvilinear coordinates  \cite{ProfBook}. Moreover, accelerometers also include an additional gravity component. Thus, the l-frame accelerometer measurements can be computed as
\begin{equation} \label{Acc local}
\boldsymbol{f}_l = \boldsymbol{R}_b^l \boldsymbol{f}_b \Delta t - (2 \boldsymbol{\Omega}_{ie}^l + \boldsymbol{\Omega}_{el}^l) \boldsymbol{v}_l \Delta t + \boldsymbol{g}_l \Delta t,
\end{equation}
where $\boldsymbol{f}_l$ is the l-frame acceleration measurements due to the motion of the vehicle, $\boldsymbol{v}_l~=~[v_e, v_n, v_u]^{\top}$ represents the vehicle's velocity in the l-frame, $\boldsymbol{\Omega}_{ie}^l$ and $\boldsymbol{\Omega}_{el}^l$ are skew-matrices constructed by $\boldsymbol{\omega}_{ie}^l$ and $\boldsymbol{\omega}_{el}^l$, respectively, and $\boldsymbol{g}_l~=~[0, 0, g]^{\top}$ is the gravitational force vector in the l-frame. 
The true vehicle accelerations $\boldsymbol{f}_l$ can be used to compute the velocities of the vehicle in the l-frame by simple discrete summation/integration, as shown in the following equation:

\begin{equation}
\boldsymbol{v}_{l_k} = \boldsymbol{v}_{l_{k-1}} + \boldsymbol{f}_{l_k}.
\end{equation}

\noindent The position of the vehicle can then be computed by integrating the velocities over time, as shown in (\ref{Velx}).

\begin{equation}\label{Velx}
\begin{split}
\varphi_k &= \varphi_{k-1} + \frac{v_{n_k}}{M + h} \Delta t\\
\lambda_k &= \lambda_{k-1} + \frac{v_{e_k}}{(N + h) \cos(\lambda_{k-1})} \Delta t\\
h_k &= h_{k-1} + v_{u_k}  \Delta t
\end{split}
\end{equation}

\subsubsection{Mechanization Stopping Mechanism}\label{Stopping Mech}
As discussed earlier, IMUs are prone to noise and biases that will accumulate over time due to their inherent design of the INS mechanization process. Yet, the mechanization process does not need to be conducted during stationary periods. Thus, if accurate knowledge of when the vehicle is stationary can be acquired, then the mechanization process can be halted during that time to avoid the accumulation of errors. Here, we propose a simple INS mechanization-stopping criteria
\begin{equation}\label{Stopping Criteria}
 \text{stationary} = 
\begin{cases}
1, & |v| \leq v_{\varepsilon} \land v_{\text{Odo}} = 0, \\
0, & \text{otherwise},
\end{cases}
\end{equation}
where `stationary' is a binary variable that indicates whether the vehicle is stationary (1) or not (0),  $|v|$ is the absolute velocity of the vehicle computed by the INS mechanization, and $v_\varepsilon$ is the velocity threshold. Thus, if the mechanized velocity is less than or equal to the threshold and the odometer reading also reports a zero velocity, then the vehicle is considered stationary; otherwise, it is not. Defining the value of $v_\varepsilon$ is critical, as choosing a lower threshold will result in more false positive detections and vice versa. Moreover, adding a condition on the odometer velocity to be strictly zero is important to guard against biased accelerometer readings that might cause false positives. Likewise, the mechanized velocity condition is needed to guard against the odometer's low resolution which does not allow for motion detection at low speeds. Additionally, odometers have much lower measurement rates compared to IMUs. Hence, relying solely on odometers to dictate the stationarity of the vehicle will prevent capturing the initial dynamics of the vehicle when it starts accelerating.

\subsection{Proposed 5G-OBMS Loosely-Coupled Integration}
In this section, we propose an LC integration between 5G mmWave BSs and onboard motion sensors via an EKF. LC integration was proposed to avoid using non-linear measurement models for the reasons explained earlier. Towards that end, we will first display the state vector and its transition model followed by the proposed measurement vector and its corresponding measurement model.

\subsubsection{NLOS Detection and switching mechanism}
As mentioned above, this paper focuses on LOS 5G positioning as an aiding solution to reset INS errors. This means that NLOS 5G measurements will bias the estimator and need to be removed. To distinguish between LOS and NLOS operation, we rely on an NLOS detection methodology proposed in \cite{bader_nlos_2022}. The method relies on the fact that time-based and power-based range measurements from NLOS sources will have a noticeable discrepancy. This is because the power-based range measurement will be affected by both the path loss (range) and power absorbed by the scattering surface while the time-based measurement is affected by the range only. In LOS scenarios, however, the discrepancy is minimal, as it is mainly caused by thermal noise. Such discrepancy is large enough to define an empirical threshold to distinguish between LOS and NLOS signals. In our implementation, the NLOS detection method is used to exclude NLOS BSs from the measurement vector. Additionally, when a total NLOS operation is detected, i.e., UE is in NLOS with all BSs, then the positioning solution will switch to a standalone INS operation mode.


\subsubsection{States and State Transition Model}
The proposed EKF filter implementation estimates the nine aforementioned PVA navigation states along with the three gyroscopes' and three accelerometers' biases. Thus, the proposed EKF has fifteen states in total, as seen in (\ref{States}). 

\begin{equation}\label{States}
    \boldsymbol{x}=\begin{bmatrix}
        &\varphi &\lambda &h &\dots\\
        &v_e &v_n &v_u &\dots\\
        &p &r &a &\dots\\
        &\delta\omega_{x} &\delta\omega_{y} &\delta\omega_{z} &\dots\\
        &\delta f_{x} &\delta f_{y} &\delta f_{z}
    \end{bmatrix}^T
\end{equation}

\noindent The process of adding sensor biases to the filter's states is known as state augmentation. This is done when state disturbances are highly correlated, e.g., the sensor biases and sensor noise. 

The transition model for the proposed LC integration primarily consists of INS mechanization described in the previous section as well as the first-order Gauss-Markov model for the sensor biases. The mechanization-based transition model, $\boldsymbol{f}(\boldsymbol{x}_{k-1}^+,\boldsymbol{u}_k)$, utlizes the accelerometer and gyroscope measurements as its inputs, i.e., $\boldsymbol{u}=[f_x,f_y,f_z,\omega_x,\omega_y,\omega_z]^{\top}$. To propagate the six sensor biases, the first-order Gauss-Markov transition model is utilized as follows
\begin{equation}\label{Gauss}
\begin{split}
        \delta \Dot{\omega}&=-\beta_\omega \delta \omega + \sqrt{2\beta_\omega}w_{\omega_B}(t),\\
        \delta \Dot{f}&=-\beta_f \delta f + \sqrt{2\beta_f}w_{f_B}(t),
\end{split}
\end{equation}
where $\beta_f$ and $\beta_\omega$ are the correlation times of the first-order Gauss-Markov models of the accelerometer and gyroscope measurements, respectively, and $w_{f_B}(t)$ and $w_{\omega_B}(t)$ are the noises that drive the biases of the accelerometers and gyroscopes, respectively. The noise components $w_{f_B}(t)$ and $w_{\omega_B}(t)$ are AWGN with variances of  $\sigma^2_{f_B}$ and $\sigma^2_{\omega_B}$, respectively. Such variances and correlation times are usually reported by sensor manufacturers and can also be estimated using Allan graph methodologies. In order to transition the covariance matrix $\boldsymbol{P}_k^-$, the Jacobian of the mechanization process and the first-order Gauss-Markov equations are computed as seen in (\ref{INS Transition}).
\begin{equation}\label{INS Transition}
\boldsymbol{\Phi} = \boldsymbol{I}_{15\times15} + \begin{bmatrix}
\Tilde{\boldsymbol{F}}_{9\times9} & \Tilde{\boldsymbol{G}}_{9\times6}\\
\boldsymbol{0}_{6\times9} & \hat{\boldsymbol{F}}_{6\times6}
\end{bmatrix}
\end{equation}
Here, the Jacobian of the mechanization process is denoted by $\Tilde{\boldsymbol{F}}$, the effect of the sensor biases and noises on the navigation states is captured by $\Tilde{\boldsymbol{G}}$, and the Jacobian of the first-order Gauss-Markov process is depicted as $\hat{\boldsymbol{F}}$. The computation of $\Tilde{\boldsymbol{F}}$ can be divided into three sub-matrices for the computation of position, velocity, and attitude states, respectively, as shown in (\ref{F}).
\begin{equation} \label{F}
    \Tilde{\boldsymbol{F}}=\begin{bmatrix}
        \Tilde{\boldsymbol{F}}_{Pos_{3\times9}} &
        \Tilde{\boldsymbol{F}}_{Vel_{3\times9}} &
        \Tilde{\boldsymbol{F}}_{Att_{3\times9}}
    \end{bmatrix}^{\top}
\end{equation}
The computation of $\Tilde{\boldsymbol{F}}_{Pos}$, $\Tilde{\boldsymbol{F}}_{Vel}$, and $\Tilde{\boldsymbol{F}}_{Att}$ are shown in \cite{ProfBook} and the computation of $\Tilde{\boldsymbol{G}}$ and $\hat{\boldsymbol{F}}$ are shown in (\ref{G tilde}) and (\ref{F hat}), respectively.
\begin{equation}\label{G tilde}
\Tilde{\boldsymbol{G}} = \begin{bmatrix}
\boldsymbol{0}_{3\times3} & \boldsymbol{0}_{3\times3} \\
\boldsymbol{0}_{3\times3} & \boldsymbol{R}^l_{b_{3\times3}}\\
\boldsymbol{R}^l_{b_{3\times3}} & \boldsymbol{0}_{3\times3}
\end{bmatrix}
\end{equation}
\begin{equation}\label{F hat}
\hat{\boldsymbol{F}} = \text{diag}\left(\begin{bmatrix}
    \boldsymbol{-\beta}_{\omega_{3\times1}} & \boldsymbol{-\beta}_{f_{3\times1}}
\end{bmatrix}\right)
\end{equation}
The processes covariance matrix $\boldsymbol{Q}$ of the proposed method comprises noises of the six sensors, as well as the noises that drive the first-order Gauss-Markov biases:
\begin{equation}\label{Q}
\boldsymbol{Q} = \text{diag}\left(\begin{bmatrix}
    \boldsymbol{\sigma}^2_{\omega_N{3\times1}} & \boldsymbol{\sigma}^2_{f_N{3\times1}} & \boldsymbol{\sigma}^2_{\omega_B{3\times1}} & \boldsymbol{\sigma}^2_{f_B{3\times1}}
\end{bmatrix}\right),
\end{equation}
where $\boldsymbol{\sigma}^2_{\omega_N{3\times1}}$ and $\boldsymbol{\sigma}^2_{f_N{3\times1}}$ are the noises of the gyroscopes and accelerometers, respectively. The proposed $\boldsymbol{Q}$ matrix differs from the ones usually proposed in the literature for both 5G/INS or GNSS/INS integration schemes. Usually, the $\boldsymbol{Q}$ matrix would encompass the noises associated with all the states  \cite{ProfBook}. From the point of view of the authors, tuning such a covariance matrix is tedious, not straightforward, and does not have scientific grounds. Tuning of such covariance matrices is either done arbitrarily, which leads to sub-optimal filtering, or through heuristic assumptions, which are not usually optimal. In contrast, the proposed $\boldsymbol{Q}$ matrix, shown in (\ref{Q}), only considers noises that are physically observable at the sensor level. Therefore, the tuning of the proposed $\boldsymbol{Q}$ matrix becomes straightforward and intuitive. 

Finally, we propose the usage of the $\boldsymbol{G}$ coupling matrix to translate the sensor noises encapsulated in the proposed $\boldsymbol{Q}$ matrix to the state covariance matrix $\boldsymbol{P}$. The proposed noise coupling matrix $\boldsymbol{G}$ has two main components, $\Tilde{\boldsymbol{G}}$ and $\hat{\boldsymbol{G}}$. The former is responsible for translating IMU noises to the PVA states while the latter is responsible for driving the IMU bias states.

\begin{equation}\label{G}
\boldsymbol{G} = \begin{bmatrix}
\Tilde{\boldsymbol{G}}_{9\times6} & \boldsymbol{0}_{9\times6}\\
\boldsymbol{0}_{6\times6} & \hat{\boldsymbol{G}}_{6\times6}
\end{bmatrix}
\end{equation}

\begin{equation}\label{G hat}
\hat{\boldsymbol{G}} = \text{diag}\left(\begin{bmatrix}
    \sqrt{2\boldsymbol{\beta}_{\omega_{3\times1}}} & \sqrt{2\boldsymbol{\beta}_{f_{3\times1}}}
\end{bmatrix}\right)
\end{equation}

\subsubsection{Measurements and Measurement Model}
In the proposed LC integration scheme, the measurement vector $\boldsymbol{z}$ comprises the 3D positioning solution of a decentralized 5G Multi-BS  integration approach, proposed in \cite{saleh_would_2022}, and the 3D velocities of the projected odometer measurements as seen in (\ref{Z}). The implementation of the proposed method is dynamic to include a varying number of available BSs. Also, it takes into account that in many cases, only a position update is available and vice versa.

\begin{equation}\label{Z}
    \boldsymbol{z}= \begin{bmatrix}
        \varphi_{5G} & \lambda_{5G} & h_{5G} &  v_{e_{\text{Odo}}} & v_{n_{\text{Odo}}} & v_{u_{\text{Odo}}}
    \end{bmatrix}^{\top}
\end{equation}

\noindent The linear measurement model of the proposed method is shown in (\ref{H INS}) and the corresponding measurement covariance matrix is shown in (\ref{R}).
\begin{equation}\label{H INS}
    \boldsymbol{H}= \begin{bmatrix}
        \boldsymbol{I}_{3\times3} & \boldsymbol{0}_{3\times3} & \boldsymbol{0}_{3\times9}\\
        \boldsymbol{0}_{3\times3} & \boldsymbol{I}_{3\times3} & \boldsymbol{0}_{3\times9}
    \end{bmatrix}
\end{equation}
\begin{equation}\label{R}
\begingroup 
\setlength\arraycolsep{0.5pt}
    \boldsymbol{R}= \text{diag}\left(\begin{bmatrix}
        \boldsymbol{\sigma}^2_{\varphi_{5G}} &\boldsymbol{\sigma}^2_{\lambda_{5G}} &\boldsymbol{\sigma}^2_{h_{5G}} &\boldsymbol{\sigma}^2_{v_{e_{\text{Odo}}}} &\boldsymbol{\sigma}^2_{v_{n_{\text{Odo}}}} &\boldsymbol{\sigma}^2_{v_{u_{\text{Odo}}}}
    \end{bmatrix}\right)
    \endgroup
\end{equation}
The position entries of the $\boldsymbol{R}$ matrix are set by the $\boldsymbol{P}^+$ postiriori covariance matrix provided by the decentralized 5G integrated solution from \cite{saleh_would_2022}. The velocity entries of the $\boldsymbol{R}$ matrix are set based on empirical tests of the odometer velocity error. 

\section{Experimental Setup}
\begin{figure}[t!]
	\centering
	\includegraphics[width=\columnwidth]{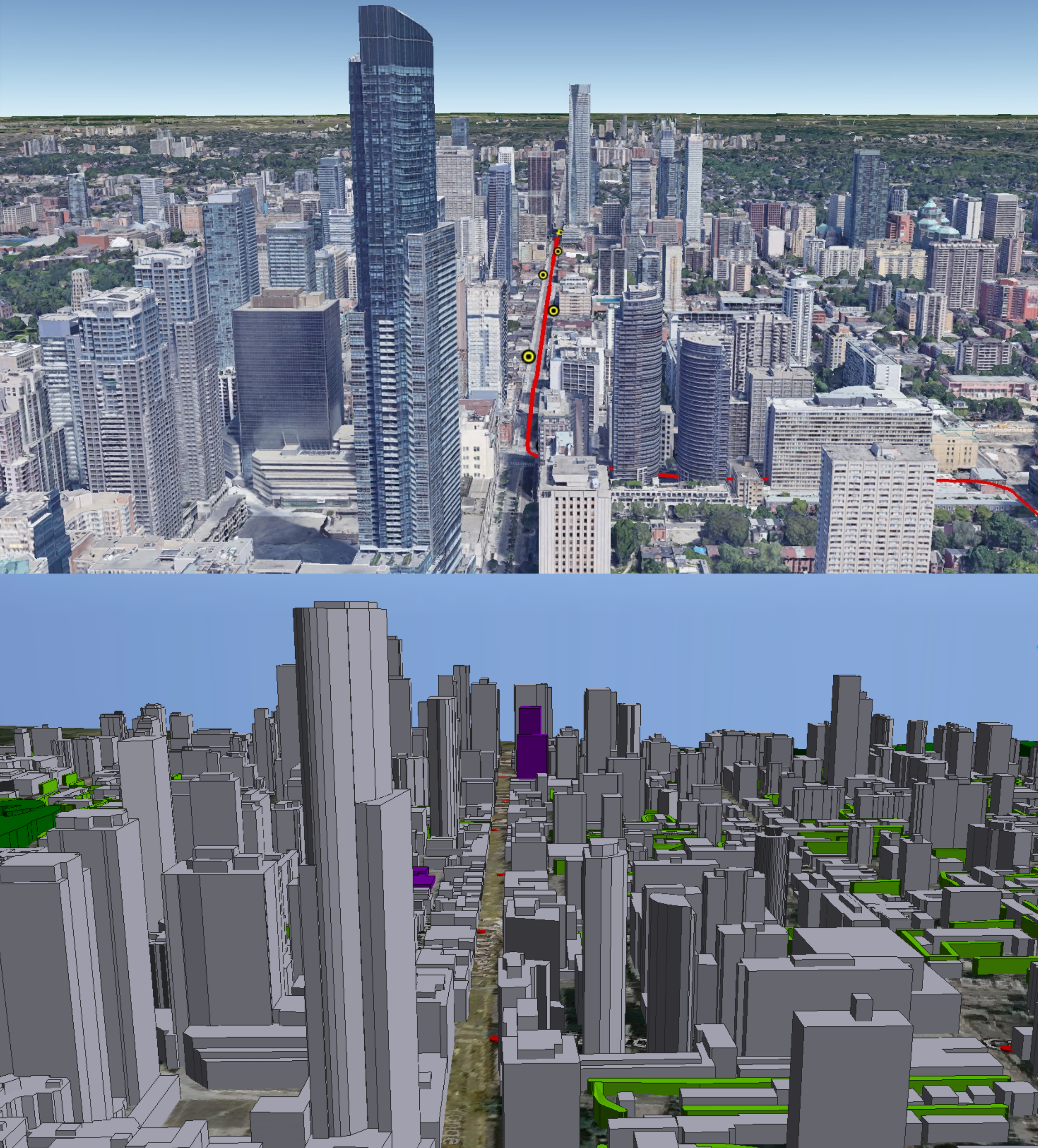}
	\DeclareGraphicsExtensions.
	\caption{Downtown Toronto, ON, Google Earth (Top) vs Siradel simulation tool (Bottom).}
	\label{GoogleEarth vs Siradel}
\end{figure}

In order to validate the proposed methods, a quasi-real simulation setup was implemented with the aid of 5G\_Channel simulator by Siradel \cite{noauthor_siradel_nodate}. Siradel's 5G\_Channel simulator is built upon LiDAR-based maps of the buildings, vegetation, and water bodies of real-world cities as seen in Fig.~\ref{GoogleEarth vs Siradel}. The simulation tool requires the position of the UE and the virtually connected BSs to compute required positioning measurables, such as RTT and AOD via its ray-tracing capabilities and propagation models. The UE and the BSs radios were configured to utilize mmWave signals with a carrier frequency of $28$ GHz and a bandwidth of $400$ MHz. The BSs were equipped with 1D $8\times1$ ULAs while the UE was equipped with an omnidirectional antenna.

To mimic a true urban navigation scenario, a vehicle was equipped with a high-end positioning solution from NovAtel's ProPak6 \cite{NovAtel}; which includes KVH-1750, a high-end tactical grade IMU, along with NovAtel's tactical grade GNSS receiver, and was driven in Downtown Toronto. NovAtel's integrated solution was then used to generate the ground truth of the trajectory performed in this work. The reference solution has a sampling frequency of $5$ Hz. The setup also included commercial grade IMUs from TPI/VTI \cite{TPI} and an odometer from OBD-II \cite{OBD} with sampling frequencies of $20$ Hz and $1$ Hz, respectively. The various sensors were connected through a ROS network to ensure synchronization between the measurements. Next, an algorithm was developed to deploy BSs along the driven trajectory, with $250$ m inter-cell distance, as per 3GPP's Release 16 guidelines \cite{Course}.

\begin{figure}[t!]
	\centering
	\includegraphics[width=\columnwidth]{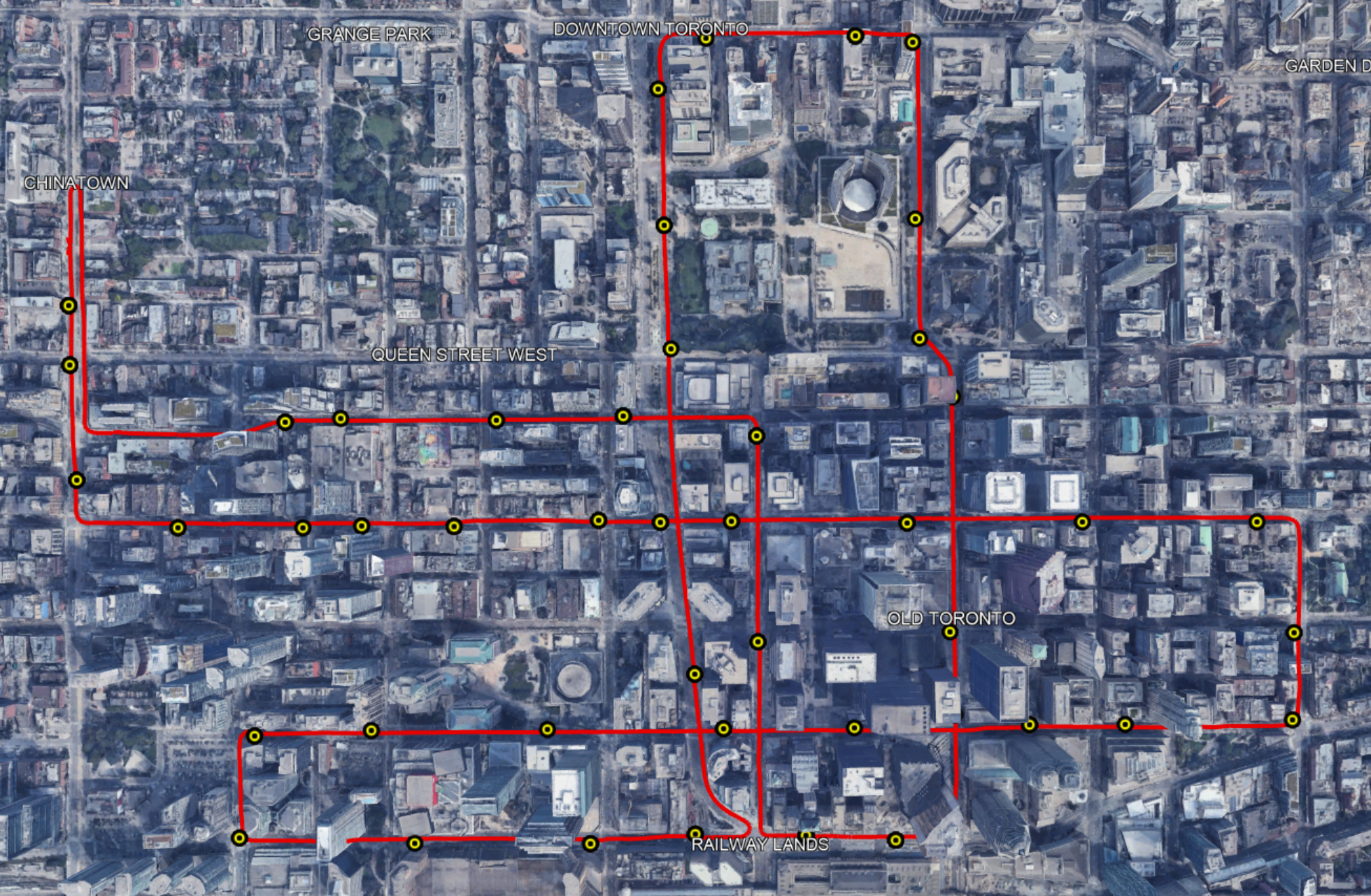}
	\DeclareGraphicsExtensions.
	\caption{Downtown Toronto Trajectory (Red), and 5G BSs (Yellow circles).}
	\label{Traj}
\end{figure}

A challenging trajectory was designed in the dense urban environment of Downtown Toronto, as seen in Fig. \ref{Traj}. The vehicle was driven for $1$ hr and $13$ mins, and the trajectory was $9$ km long. The trajectory was conducted during rush hour, and thus, many sudden car stopping/acceleration dynamics were encountered, challenging the OBMS solution. The trajectory also featured numerous turning dynamics that caused several natural total 5G outages. To be specific, four natural 5G outages were encountered which accumulated to $3\%$ of the trajectory. These four outages varied in duration which resulted in varying challenging scenarios for the proposed solution. The first outage is $13$ seconds long, the second outage is $8$ seconds long, the third outage is $20$ seconds long, and the fourth outage is $100$ seconds long. For the rest of the trajectory, the UE was simultaneously connected to three BSs for $30\%$ of the time, two BSs for $46\%$ of the time, and a single BS for $21\%$ of the time.


\section{Results and Discussion}
In this section, positioning performance results will be shown for the proposed initial bias removal approach, stationary mechanization stopping mechanism, and the LC integration between 5G and OBMS.

\begin{figure}[t!]
	\centering
	\includegraphics[trim=117pt 230 125pt 245pt,clip,width=\columnwidth]{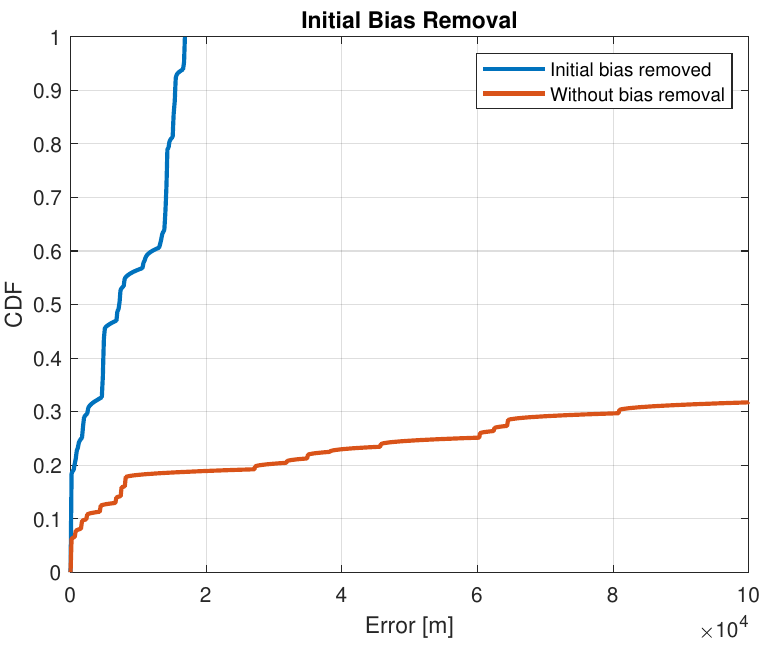}
	\DeclareGraphicsExtensions.
	\caption{CDF of the positioning errors for standalone INS implementations with/without initial bias removal.}
	\label{CDF Bias}
\end{figure}
\begin{table}[t!]
	\caption{3D Positioning Error Statistics for INS Implementations With/Without Initial Bias Removal}
	\label{INS Bias Errors}
	\begin{tabularx}{\columnwidth}{@{}l*{2}{C}c@{}}
		\toprule
		&Statistics &Bias removed 	   &W/o bias removal\\
		\midrule
		&RMS        &10 km    &430 km\\ 
		&Max        &17 km    &630 km\\
		&$<2$ m     &5.9\%    &5.8\%\\ 
		&$<1$ m     &5.8\%    &5.8\%\\
		&$<30$ cm   &5.7\%    &5.4\%\\
		\bottomrule
	\end{tabularx}
\end{table}

\subsection{Initial Bias Removal}
In order to gauge the importance of initial bias removal, a standalone INS mechanization was done on raw IMU data with and without initial bias removal. The CDF of the 3D positioning error is shown in Fig. \ref{CDF Bias}. It can be seen that both solutions sustained a high amount of error, as they both accumulated biases for more than an hour. Yet, it can be seen that when initial biases were removed, the accumulation of error occurs at a much slower pace compared to mechanizing without bias removal. Table \ref{INS Bias Errors} showcases the error statistics for both implementations. It can be seen that both implementations could not sustain a sub-2 meter level of accuracy for more than $6\%$ of the time. Yet, the maximum error of the solution with bias removal was capped compared to its counterpart. Therefore, the results of the next experiments will all include initial bias removal using the methodology highlighted above.

\subsection{Stationary Stopping Mechanism}
\begin{figure}[t!]
	\centering
	\includegraphics[trim=117pt 230 125pt 245pt,clip,width=\columnwidth]{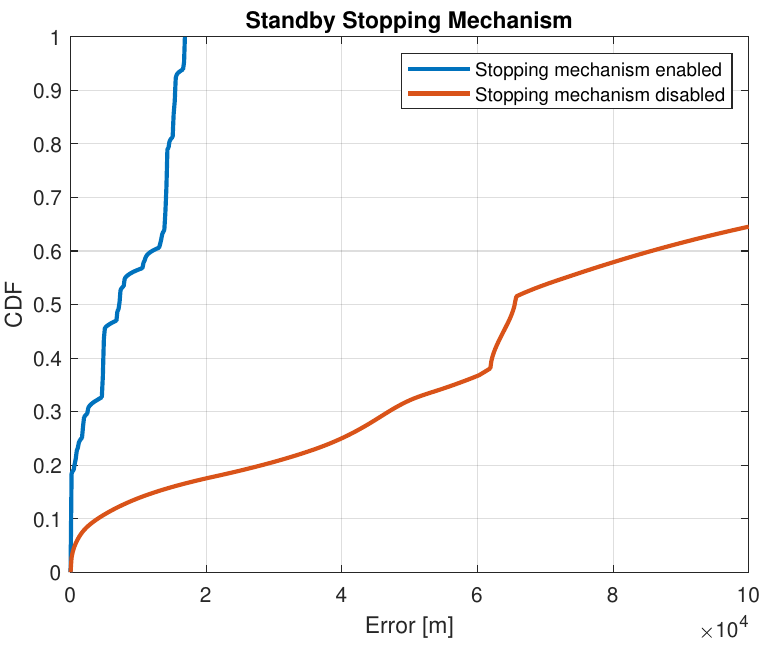}
	\DeclareGraphicsExtensions.
	\caption{CDF of the positioning errors for standalone INS implementations with/without mechanization stopping mechanism while not in motion.}
	\label{Stopping CDF}\vspace{-2pt}
\end{figure}

In order to validate the need for a stopping mechanism while the vehicle is in stationary mode, two INS mechanization solutions were tested with and without the proposed stopping mechanism. Fig. \ref{Stopping CDF} shows the CDF of the 3D positioning error for both solutions. As is the case with previous results, positioning by utilizing a standalone IMU solution for long periods of time is not sustainable. Yet, it can be seen that even with initial bias removal for both implementations, there would be residual IMU biases. These biases caused the solution without a stopping mechanism to accumulate more errors during stationary periods as seen in the figure. The error statistics of both implementations are shown in Table \ref{INS Stopping Errors}. It can be seen that the stopping mechanism enhanced all error statistics by an order of magnitude. Such an improvement is crucial while integrating with 5G and targeting a deci-meter level of accuracy for AV applications.

\begin{table}[t!]
	\caption{3D Positioning Error Statistics for INS Implementations With/Without Mechanization Stopping Mechanism}
	\label{INS Stopping Errors}
	\begin{tabularx}{\columnwidth}{@{}l*{2}{C}c@{}}
		\toprule
		&Statistics &Stopping mech. 	 &W/o stopping mech.\\
		\midrule
		&RMS        &10 km    &100 km\\ 
		&Max        &17 km    &190 km\\
		&$<2$ m     &5.9\%    &0.49\%\\ 
		&$<1$ m     &5.8\%    &0.37\%\\
		&$<30$ cm   &5.7\%    &0.21\%\\
		\bottomrule
	\end{tabularx}
\end{table}

\subsection{INS Integration with Odometer}
\begin{figure}[t!]
	\centering
	\includegraphics[trim=117pt 230 125pt 245pt,clip,width=\columnwidth]{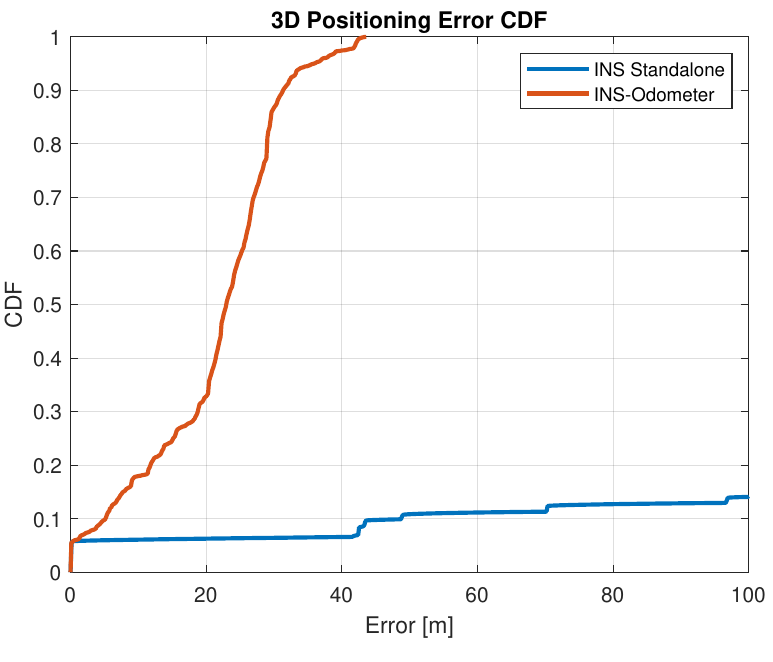}
	\DeclareGraphicsExtensions.
	\caption{CDF of the positioning errors for standalone INS and INS-odometer integration.}
	\label{INS-Odo CDF}
\end{figure}

Before testing the proposed LC integration between 5G and OBMS, it is worth examining the effect of the integration between INS and odometer measurements. This is important as the methods in the literature omitted integration with odometers, despite their availability in all modern vehicles. Thus, the standalone INS mechanization will be compared to INS-Odometer EKF integration. Fig. \ref{INS-Odo CDF} shows the CDF of the 3D positioning error for both solutions. It can be seen that integration with odometer measurements has capped the maximum error to around $40$ m down from $17$ km. This is due to the fact that velocity updates would reset acceleration errors when available, limiting the accumulation of bias. The error statistics of both solutions are shown in Table \ref{INS-Odo Errors} for more insights. It can be seen that despite being successful in limiting the maximum and RMS errors of standalone INS, integration with odometer measurements was not enough to enhance the sub-meter error statistics significantly. This is mainly attributed to two reasons. First, odometer updates do not directly reset gyroscope biases. Second, as the odometer update rate is fixed to $1$ Hz while  the IMU runs at $20$ Hz, the IMU will inevitably accumulate positioning errors between the velocity updates. These errors will not be totally reset by velocity measurements, as they only have observability over velocity and acceleration states.

\begin{table}[t!]
	\caption{3D Positioning Error Statistics for INS Standalone and INS-Odometer Integration}
	\label{INS-Odo Errors}
	\begin{tabularx}{\columnwidth}{@{}l*{2}{C}c@{}}
		\toprule
		&Statistics  &INS SA 	&INS-Odo.\\
		\midrule
		&RMS         &10 km     &23.5 m\\ 
		&Max         &17 km     &43.6 m\\
		&$<2$ m      &5.9\%     &7.1\%\\ 
		&$<1$ m      &5.8\%     &6.1\%\\
		&$<30$ cm    &5.7\%     &5.6\%\\
		\bottomrule
	\end{tabularx}
\end{table}

\subsection{LC 5G-OBMS Integration}
To benchmark our proposed LC 5G-OBMS integration, we can either use a 5G-OBMS TC scheme or a 5G-constant velocity model LC scheme as a benchmark. In a prior investigation, documented in \cite{saleh_5g-enabled_2022}, we compared LC integration of multiple BSs to TC integration while using a constant velocity model for each method. The results in \cite{saleh_5g-enabled_2022} show that the LC scheme outperformed its TC counterpart, which was primarily attributed to the substantial linearization errors inherent in the TC's measurement model. Drawing from these conclusions, we recognize that changing the transition model (i.e., using IMU) will not rectify the limitations of the TC's measurement model. Consequently, we anticipate that benchmarking our proposed method with a 5G-OBMS TC will lead to the same conclusions highlighted in our previous work. Thus, a better study would be to assess the incremental performance enhancement facilitated by IMU integration compared to a constant velocity model while using an LC scheme for both methods.
\begin{figure}[t!]
	\centering
	\includegraphics[trim=117pt 243 125pt 245pt,clip,width=\columnwidth]{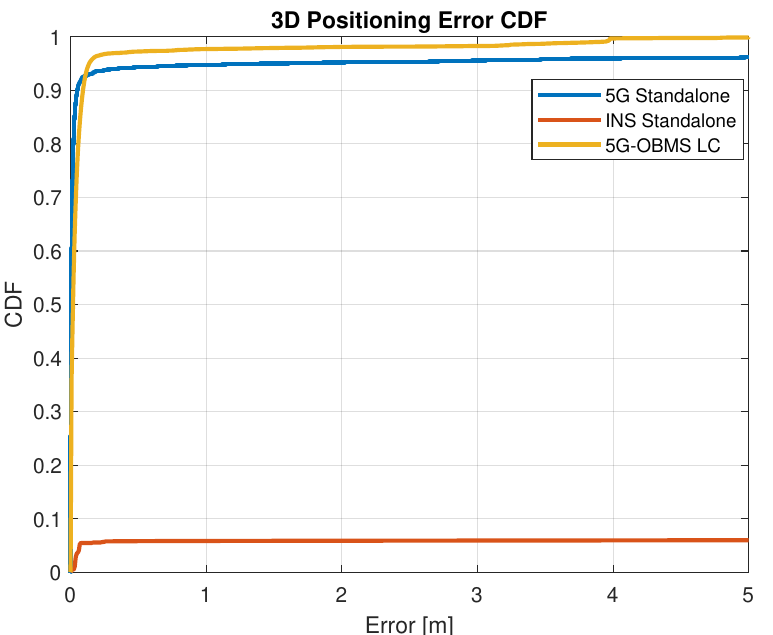}
	\DeclareGraphicsExtensions.
	\caption{CDF of the positioning errors while utilizing a standalone 5G hybrid multi-BS solution, standalone INS solution, and the proposed LC 5G-OBMS integration solution.}
	\label{Integration CDF}
\end{figure}
\begin{table}[t!]
	\caption{3D Positioning Error Statistics For 5G Standalone, INS Standalone, and 5G-OBMS}
	\label{INS vs 5G}
    \vspace{-2pt}
	\begin{tabularx}{\columnwidth}{@{}l*{3}{C}c@{}}
        \toprule
		&Statistics  &5G-SA 	&INS-SA    &5G-OBMS\\
		\midrule
		&RMS         & 9 m      &10 km     &0.5 m\\ 
		&Max         & 89.3 m   &17 km     &6.3 m\\
		&$<2$ m      & 95.2\% 	&5.9\%     &98.1\%\\ 
		&$<1$ m      & 94.7\% 	&5.8\%     &97.7\%\\
		&$<30$ cm    & 93.9\% 	&5.7\%     &96.9\%\\
		\bottomrule
	\end{tabularx}
\end{table}
Fig. \ref{Integration CDF} shows the 3D positioning error CDFs for 5G standalone, INS standalone, and 5G-OBMS LC integration solutions. As expected, the INS standalone solution using commercial-grade IMUs could not sustain a sub-meter level of accuracy for more than $6\%$ of the time, as outlined above. A more noteworthy comparison would be between the standalone 5G and 5G-OBMS LC integration. It can be seen that the proposed LC 5G-OBMS integration enhanced the decimeter level of accuracy of the 5G standalone solution. Table \ref{INS vs 5G} shows a summary of the positioning error statistics of the three solutions. It can be seen that not only did the integration with OBMS enhance the sub-meter/decimeter error statistics but also significantly decreased the maximum and RMS errors of the 5G standalone solution. The maximum error due to the constant velocity model was decreased from around $90$ m of error to $6.3$ m, which is an enhancement of $90\%$. Moreover, the RMS error statistic was significantly decreased from $9$ m to $0.5$ m, which is a $94\%$ error reduction. It is worth noting that the proposed method has a $2\sigma$ error statistic of $14$ cm. This means that the solution can sustain errors below $14$ cm for $95\%$ of the time, which achieves the positioning requirements for level four of autonomous driving \cite{AVRequirements}.

\begin{figure}[t!]
	\centering
	\includegraphics[width=\columnwidth]{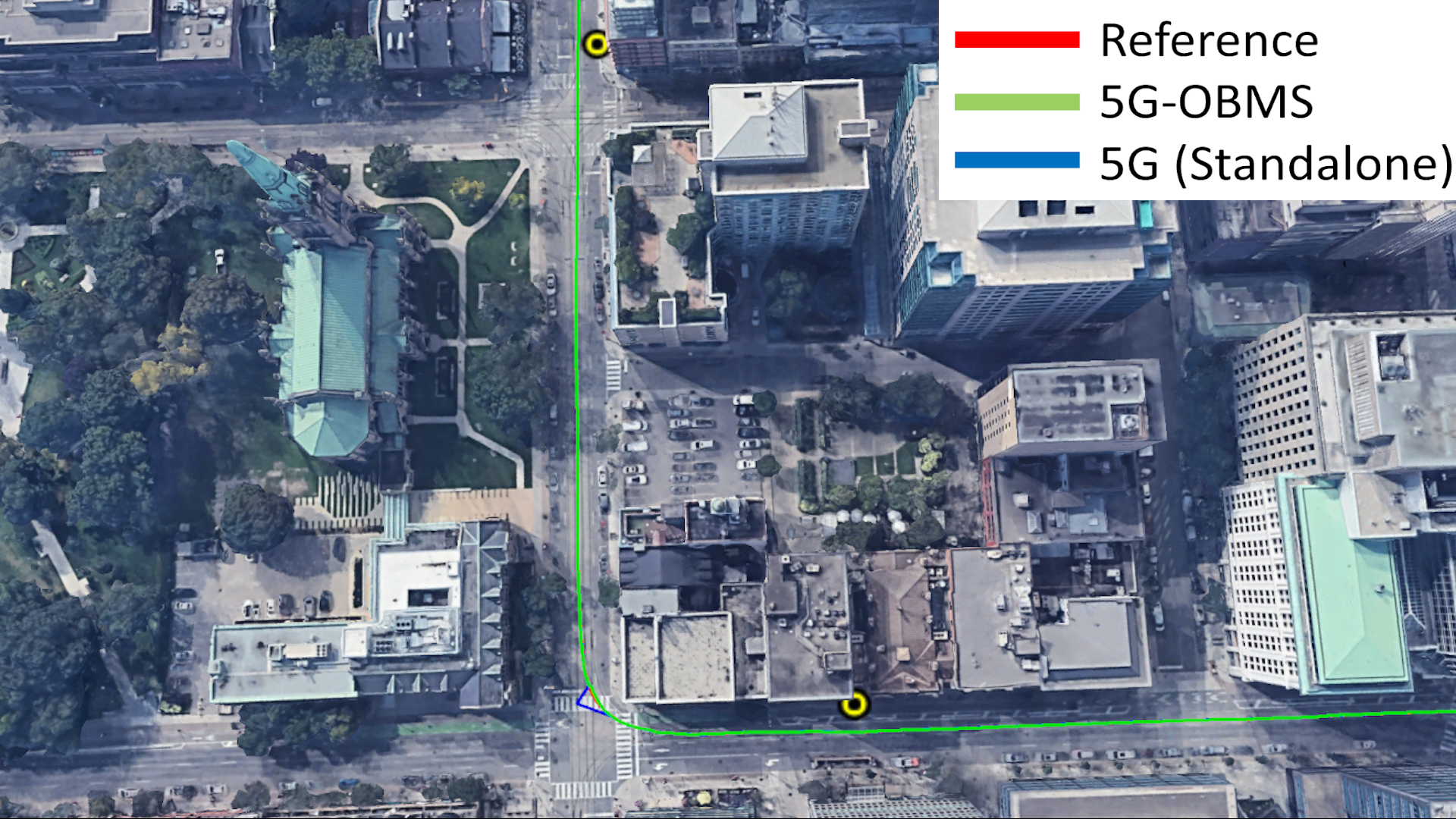}
	\DeclareGraphicsExtensions.
	\caption{First close-up to showcase the performance of 5G-OBMS integration against standalone 5G in a natural 5G outage.}
	\label{5G-INS1}
\end{figure}

\begin{figure}[t!]
	\centering
	\includegraphics[width=\columnwidth]{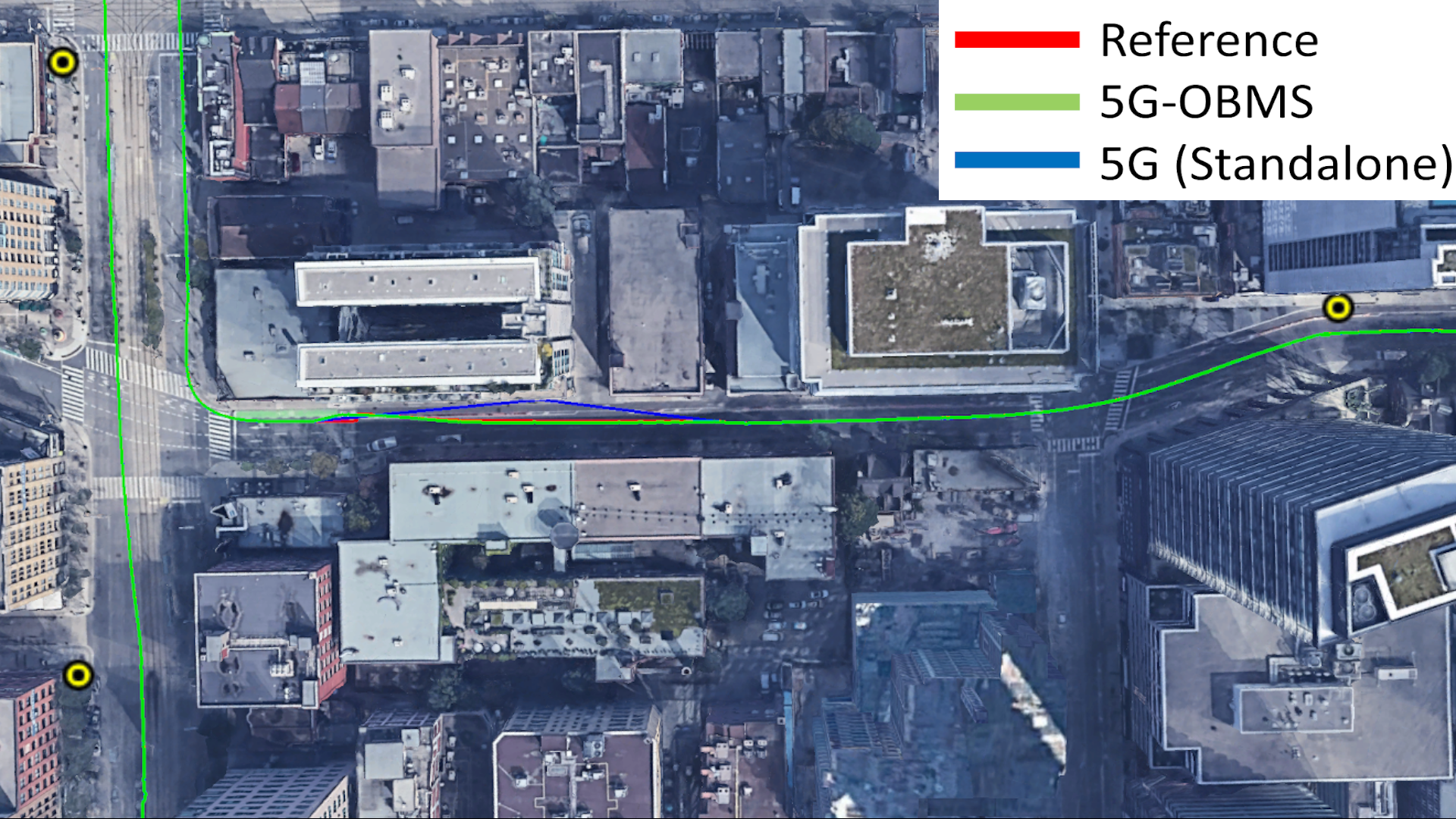}
	\DeclareGraphicsExtensions.
	\caption{Second close-up to showcase the performance of 5G-OBMS integration against standalone 5G in a natural 5G outage.}
	\label{5G-INS2}
\end{figure}

\noindent To have more insights on the proposed methodology, four close-up scenarios are shown during the four natural 5G outages, shown in Figs. \ref{5G-INS1}-\ref{5G-INS4}. 


\subsection{Close-up Scenarios}
\subsubsection{First Scenario}
In the first scenario, shown in Fig. \ref{5G-INS1}, the UE was initially connected to the bottom BS until it reached a right-turn corner, at which it experienced a short outage of around eight seconds. It can be seen that the moment of the outage occurred during a right turn dynamic, thus, the constant velocity model headed in a tangential path to the true direction of motion. It is worth noting that it is expected that such a type of outage will be the most dominant, as corners are usually not totally covered with LOS communications. On the other hand, the IMUs and odometer were capable of handling such a short outage, as the IMU biases were adequately removed.

\subsubsection{Second Scenario}
In the second scenario, shown in Fig. \ref{5G-INS2}, the vehicle was driving from the east BS toward the west direction, and then performed a right turn dynamic. It can be seen that a natural total 5G outage was encountered before the right turn, as all three BSs were blocked by buildings surrounding the UE for eight seconds. This outage occurred right after an 'S' shaped dynamic, which made the constant velocity model of the standalone 5G solution overshoot for the next few epochs. As the outage ensued, the constant velocity model was not stable enough and headed in the north direction. On the other hand, the proposed LC 5G-OBMS solution was able to bridge such a gap with ease until the UE was able to reconnect to BSs at the end of the road.

\begin{figure}[t!]
	\centering
	\includegraphics[width=\columnwidth]{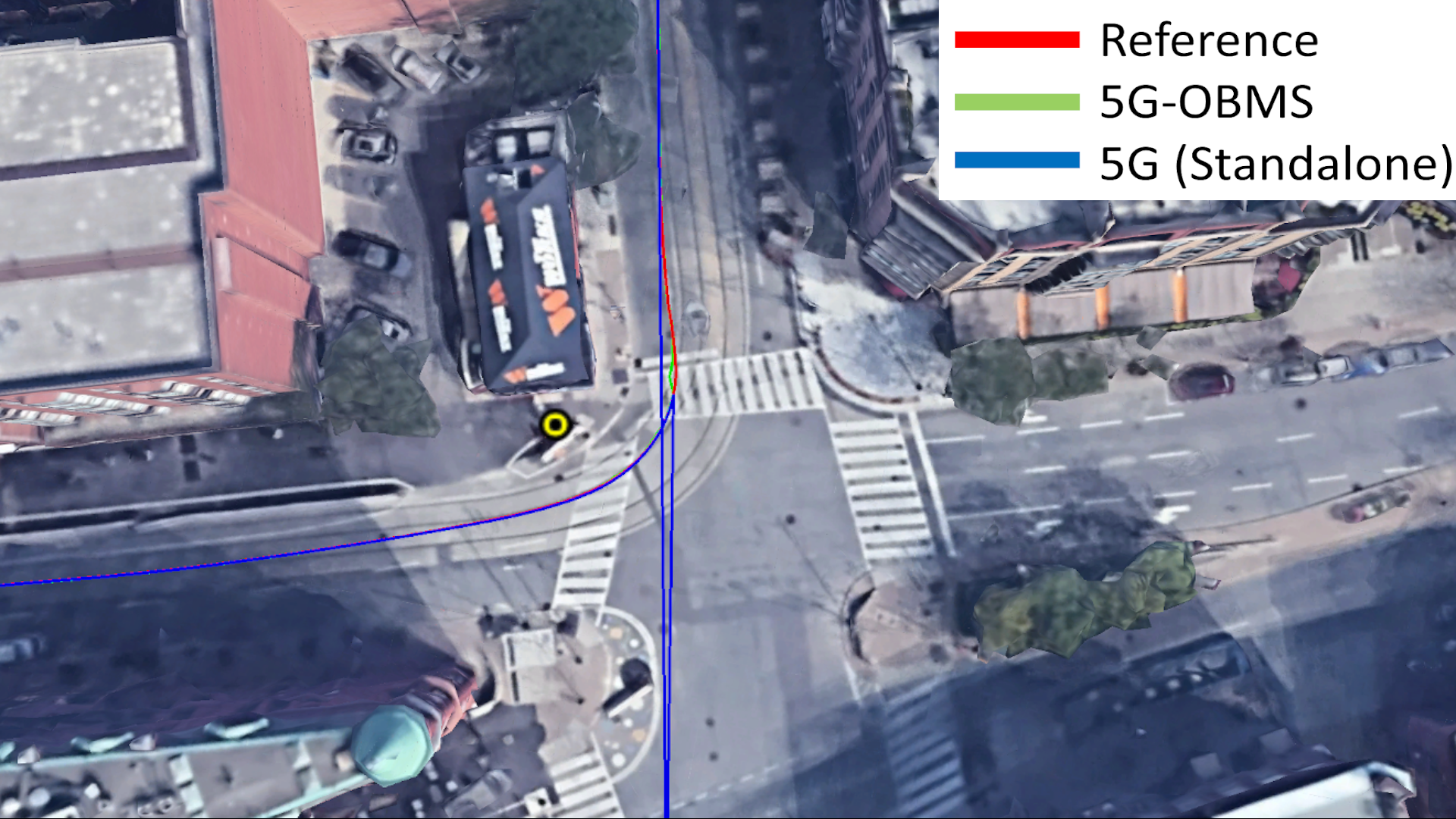}
	\DeclareGraphicsExtensions.
	\caption{Third close-up to showcase the performance of 5G-OBMS integration against standalone 5G in a natural 5G outage.}
	\label{5G-INS3}
\end{figure}
\begin{figure}[t!]
	\centering
	\includegraphics[width=\columnwidth]{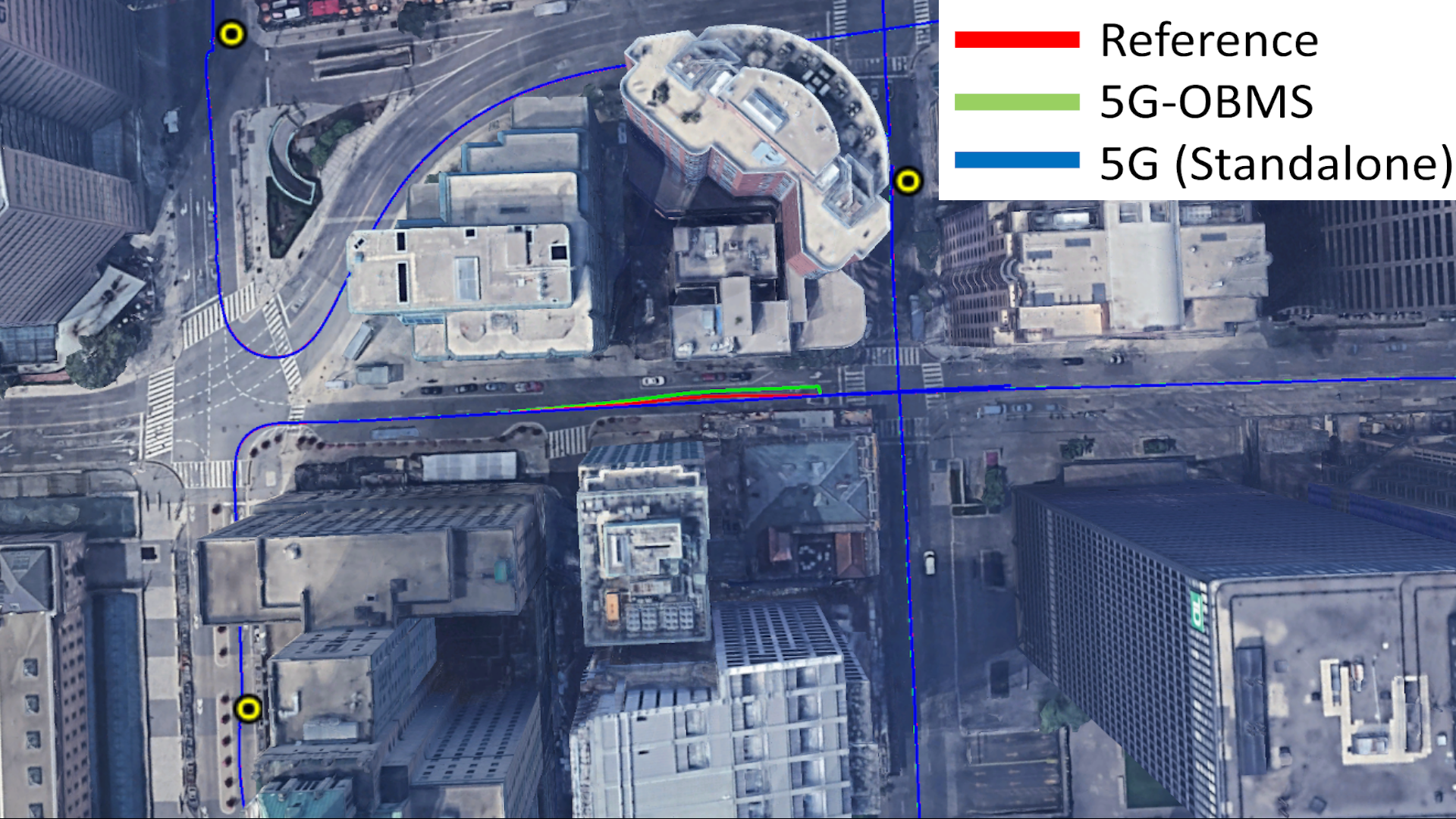}
	\DeclareGraphicsExtensions.
	\caption{Fourth close-up to showcase the performance of 5G-OBMS integration against standalone 5G in a natural 5G outage.}
	\label{5G-INS4}
\end{figure}

\begin{figure}[t!]
	\centering
	\includegraphics[trim=117pt 230 125pt 260pt,clip,width=1\columnwidth]{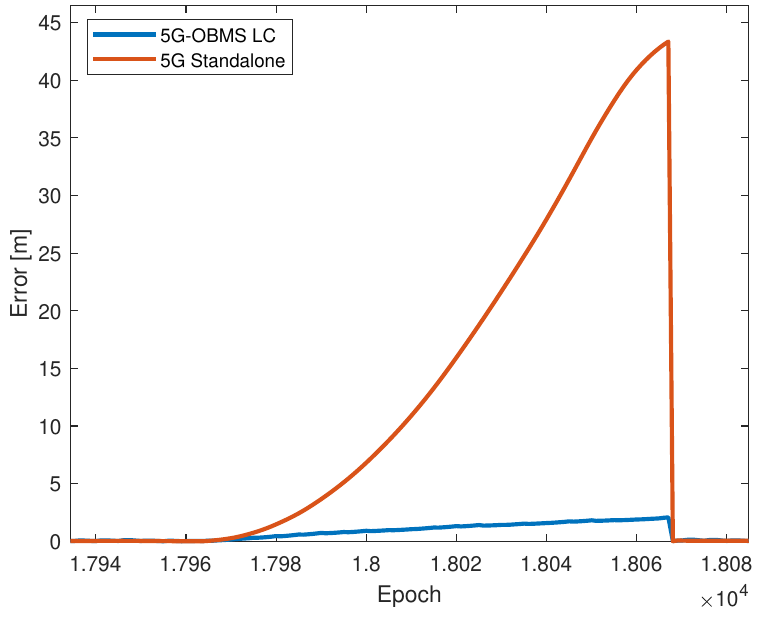}\vspace{-20pt}
	\DeclareGraphicsExtensions.
	\caption{A close-up of the error profile of the proposed 5G-OBMS method (blue) and the standalone 5G constant velocity model (red) during the fourth scenario.}
	\label{Error Closeup}
\end{figure}

\subsubsection{Third Scenario}
In the third scenario, shown in Fig. \ref{5G-INS3}, the UE was driven south as it disconnected from a BS to the north. As the next available BS was obstructed by a building, the UE experienced a 20-second outage until a right turn was conducted. This scenario resembles the natural 5G outage that was shown in the first scenario. As the outage was detected before the right-turn dynamic, the 5G standalone constant velocity model overshot in a straight line, which resulted in high amounts of error. On the other hand, the onboard motion sensors were successfully able to mitigate the natural outage, as their biases were properly reset in the previous epochs.

\subsubsection{Fourth Scenario}
The fourth scenario, shown in Fig. \ref{5G-INS4}, resembles one of the most challenging scenarios that 5G mmWave positioning can encounter in a dense urban environment. The vehicle was driving toward the north (bottom left side of Fig. \ref{5G-INS4} and performed a right turn. It can be seen that at that time, there were three BSs in the vicinity of the UE, yet, none of them were in LOS, as they were obstructed by buildings. At first glance, the reader might notice that the standalone 5G solution's constant velocity model was able to outperform the 5G-OBMS integration. Yet, a closer look at the error timeline for both solutions, shown in Fig. \ref{Error Closeup}, paints a different story. It can be clearly seen that although the proposed method has sustained a maximum error of $2$ m, yet, it has evidently outperformed the constant velocity model, which incurred around $40$ m of error over the 100-second long outage. It is worth noting that the standalone 5G error was primarily dominated by longitudinal error while the proposed 5G-OBMS method's error was primarily dominated by lateral error. That is why it is hard to distinguish the standalone 5G error in a 2D map plot. A summary of the RMS and maximum errors endured by both implementations in the four natural outage scenarios are presented in Table \ref{Four closeups}. Such statistics are important, as they isolate the effect of the proposed 5G-OBMS integration. It is clear beyond doubt that the proposed LC 5G-OBMS integration has outperformed the constant velocity model as it significantly reduced both the RMS and maximum errors during 5G outages. Additionally, the sub-meter and decimeter errors were enhanced to meet the stringent needs of higher levels of vehicular autonomy.

\vspace{-3pt}
\begin{table}[t!]
	\caption{3D Positioning Error Statistics Summary of The Four Scenarios}
	\label{Four closeups}
	\begin{tabularx}{\columnwidth}{@{}l*{5}{C}c@{}}
		\toprule
		&Statistics &First 	    &Second    &Third     &Fourth\\
		\midrule
		\multicolumn{5}{c}{\hspace{50pt} 5G SA-Constant Velocity Model}\\
		\midrule
		&RMS        & 22 m     &55 m    &18.9 m  &22.4 m\\ 
		&Max        & 39.6 m   &89.9 m  &37.2 m  &43.4 m\\
        \midrule
		\multicolumn{5}{c}{\hspace{50pt}Proposed LC 5G-OBMS Integration}\\
		\midrule
		&RMS        & 0.5 m   &3.7 m &2.2 m  &1.3 m\\ 
		&Max        & 1.2 m   &6.3 m  &2.9 m  &2.1 m\\
		\bottomrule
	\end{tabularx}
\end{table}

\vspace{-9pt}

\section{Conclusion}
5G is anticipated to provide accurate positioning estimates in GNSS-denied environments. Yet, LOS operation cannot be guaranteed in such environments. During LOS outages, standalone 5G positioning solutions will incur high positioning errors. Hence, we proposed to bridge these gaps via the integration with IMUs and odometers. The proposed sensor fusion scheme utilizes an EKF in a loosely coupled approach to avoid linearization errors arising from its tightly coupled counterpart. Additionally, we proposed a new method to design the filter's process covariance matrix that makes the tuning process easier. Moreover, we proposed minor modifications to the traditional INS mechanization where we added initial bias removal, a stationary stopping mechanism, and integration with odometers. The proposed methods were validated using real IMU measurements accompanied by quasi-real 5G measurements from a raytracing tool. The measurements were obtained from an hour-long trajectory in Downtown Toronto, during which, four natural 5G outages occurred. The proposed method was benchmarked against a loosely coupled standalone 5G solution that relied on a constant velocity model. We found that the proposed IMU enhancements and integration with an odometer are paramount to achieving high-precision positioning during 5G outages, especially while using commercial-grade IMUs. Additionally, accurate NLOS detection algorithms are required to optimally switch between 5G-OBMS and standalone OBMS operation during outages. Finally, Utilizing onboard motion sensors as a transition model outperforms the constant velocity model employed by standalone 5G methods, achieving $14$ cm accuracy for $95\%$ of the time.

\bibliographystyle{IEEEtran}
\bibliography{References,references}

\begin{thebibliography}{10}
\providecommand{\url}[1]{#1}
\csname url@samestyle\endcsname
\providecommand{\newblock}{\relax}
\providecommand{\bibinfo}[2]{#2}
\providecommand{\BIBentrySTDinterwordspacing}{\spaceskip=0pt\relax}
\providecommand{\BIBentryALTinterwordstretchfactor}{4}
\providecommand{\BIBentryALTinterwordspacing}{\spaceskip=\fontdimen2\font plus
\BIBentryALTinterwordstretchfactor\fontdimen3\font minus \fontdimen4\font\relax}
\providecommand{\BIBforeignlanguage}[2]{{%
\expandafter\ifx\csname l@#1\endcsname\relax
\typeout{** WARNING: IEEEtran.bst: No hyphenation pattern has been}%
\typeout{** loaded for the language `#1'. Using the pattern for}%
\typeout{** the default language instead.}%
\else
\language=\csname l@#1\endcsname
\fi
#2}}
\providecommand{\BIBdecl}{\relax}
\BIBdecl

\bibitem{AVCongestion}
Policy, P.~S. C. P.~W. group~on Automated, and C.~Vehicles, ``Automated and connected vehicles policy framework for canada,'' PPSC, Tech. Rep., 2019.

\bibitem{guerrero-ibanez_sensor_2018}
\BIBentryALTinterwordspacing
J.~Guerrero-Ibáñez, S.~Zeadally, and J.~Contreras-Castillo, ``\BIBforeignlanguage{en}{Sensor {Technologies} for {Intelligent} {Transportation} {Systems}},'' \emph{\BIBforeignlanguage{en}{Sensors}}, vol.~18, no.~4, p. 1212, Apr. 2018, number: 4 Publisher: Multidisciplinary Digital Publishing Institute. [Online]. Available: \url{https://www.mdpi.com/1424-8220/18/4/1212}
\BIBentrySTDinterwordspacing

\bibitem{SAE}
{SAE}, ``J3016 automated-driving graphic update,'' \url{https://www.sae.org/news/2019/01/sae-updates-j3016-automated-driving-graphic}, 2019.

\bibitem{AVRequirements}
T.~Reid, S.~Houts, R.~Cammarata, G.~Mills, S.~Agarwal, A.~Vora, and G.~Pandey, ``Localization requirements for autonomous vehicles,'' \emph{SAE International Journal of Connected and Automated Vehicles}, vol.~2, pp. 173--190, 09 2019.

\bibitem{yeong_sensor_2021}
\BIBentryALTinterwordspacing
D.~J. Yeong, G.~Velasco-Hernandez, J.~Barry, and J.~Walsh, ``\BIBforeignlanguage{en}{Sensor and {Sensor} {Fusion} {Technology} in {Autonomous} {Vehicles}: {A} {Review}},'' \emph{\BIBforeignlanguage{en}{Sensors}}, vol.~21, no.~6, p. 2140, Jan. 2021, number: 6 Publisher: Multidisciplinary Digital Publishing Institute. [Online]. Available: \url{https://www.mdpi.com/1424-8220/21/6/2140}
\BIBentrySTDinterwordspacing

\bibitem{ProfBook}
A.~Noureldin, T.~B. Karamat, and J.~Georgy, \emph{Fundamentals of Inertial Navigation, Satellite-based Positioning and their Integration}.\hskip 1em plus 0.5em minus 0.4em\relax Berlin, Heidelberg: Springer, 2013.

\bibitem{Camera}
N.~Yang, R.~Wang, X.~Gao, and D.~Cremers, ``{Challenges in monocular visual odometry: Photometric calibration, motion bias, and rolling shutter effect},'' \emph{IEEE Robotics and Automation Letters}, vol.~3, no.~4, pp. 2878--2885, 2018.

\bibitem{Radar}
A.~Abosekeen, T.~B. Karamat, A.~Noureldin, and M.~J. Korenberg, ``{Adaptive cruise control radar-based positioning in GNSS challenging environment},'' \emph{IET Radar, Sonar and Navigation}, vol.~13, no.~10, pp. 1666--1677, 2019.

\bibitem{wymeersch_integration_2021}
H.~Wymeersch, D.~Shrestha, C.~M. de~Lima, V.~Yajnanarayana, B.~Richerzhagen, M.~F. Keskin, K.~Schindhelm, A.~Ramirez, A.~Wolfgang, M.~F. de~Guzman, K.~Haneda, T.~Svensson, R.~Baldemair, and S.~Parkvall, ``Integration of {Communication} and {Sensing} in {6G}: a {Joint} {Industrial} and {Academic} {Perspective},'' in \emph{{IEEE} {Annual} {International} {Symposium} on {Personal}, {Indoor} and {Mobile} {Radio} {Communications} ({PIMRC})}, Sep. 2021, pp. 1--7.

\bibitem{merits}
S.~Dwivedi, R.~Shreevastav, F.~Munier, J.~Nygren, I.~Siomina, Y.~Lyazidi, D.~Shrestha, G.~Lindmark, P.~Ernstr{\"o}m, E.~Stare, S.~M. Razavi, S.~Muruganathan, G.~Masini, {\AA}.~Busin, and F.~Gunnarsson, ``Positioning in {5G} networks,'' \emph{IEEE Communications Magazine}, vol.~59, no.~11, pp. 38--44, 2021.

\bibitem{mogyorosi_positioning_2022}
F.~Mogyorósi, P.~Revisnyei, A.~Pašić, Z.~Papp, I.~Törös, P.~Varga, and A.~Pašić, ``Positioning in {5G} and {6G} {Networks}—{A} {Survey},'' \emph{Sensors}, vol.~22, no.~13, p. 4757, Jan. 2022.

\bibitem{Fokin2021}
G.~Fokin and A.~Vladyko, ``Vehicles tracking in {5G}-{V2X} {UDN} using range, bearing and inertial measurements,'' in \emph{2021 13th International Congress on Ultra Modern Telecommunications and Control Systems and Workshops (ICUMT)}, 2021, pp. 137--142.

\bibitem{saleh_vehicular_2021}
S.~Saleh, S.~Sorour, and A.~Noureldin, ``Vehicular {Positioning} {Using} {mmWave} {TDOA} with a {Dynamically} {Tuned} {Covariance} {Matrix},'' in \emph{{IEEE} {Globecom} {Workshops} ({GC} {Wkshps})}, Dec. 2021, pp. 1--6.

\bibitem{wen2024high}
T.~Wen, H.~Jiang, B.~Cai, and C.~Roberts, ``High-speed train positioning using improved extended kalman filter with 5g nr signals,'' \emph{IEEE Transactions on Intelligent Transportation Systems}, 2024.

\bibitem{koivisto_high-efficiency_2017}
\BIBentryALTinterwordspacing
M.~Koivisto, A.~Hakkarainen, M.~Costa, P.~Kela, K.~Leppanen, and M.~Valkama, ``\BIBforeignlanguage{en}{High-{Efficiency} {Device} {Positioning} and {Location}-{Aware} {Communications} in {Dense} {5G} {Networks}},'' \emph{\BIBforeignlanguage{en}{IEEE Communications Magazine}}, vol.~55, no.~8, pp. 188--195, Aug. 2017. [Online]. Available: \url{https://ieeexplore.ieee.org/document/7984759/}
\BIBentrySTDinterwordspacing

\bibitem{mostafavi_vehicular_2020}
S.~S. Mostafavi, S.~Sorrentino, M.~B. Guldogan, and G.~Fodor, ``Vehicular {Positioning} {Using} {5G} {Millimeter} {Wave} and {Sensor} {Fusion} in {Highway} {Scenarios},'' in \emph{{IEEE} {International} {Conference} on {Communications} ({ICC})}, Jun. 2020, pp. 1--7, iSSN: 1938-1883.

\bibitem{luo_research_2021}
Y.~Luo, M.~Wang, C.~Guo, and W.~Guo, ``\BIBforeignlanguage{en}{Research on {Invariant} {Extended} {Kalman} {Filter} {Based} {5G}/{SINS} {Integrated} {Navigation} {Simulation}},'' in \emph{\BIBforeignlanguage{en}{China {Satellite} {Navigation} {Conference} ({CSNC})}}, ser. Lecture {Notes} in {Electrical} {Engineering}, C.~Yang and J.~Xie, Eds.\hskip 1em plus 0.5em minus 0.4em\relax Singapore: Springer, 2021, pp. 455--466.

\bibitem{wang_simulation_2022}
\BIBentryALTinterwordspacing
Y.~Wang, B.~Zhao, W.~Zhang, and K.~Li, ``\BIBforeignlanguage{en}{Simulation {Experiment} and {Analysis} of {GNSS}/{INS}/{LEO}/{5G} {Integrated} {Navigation} {Based} on {Federated} {Filtering} {Algorithm}},'' \emph{\BIBforeignlanguage{en}{Sensors}}, vol.~22, no.~2, p. 550, Jan. 2022, number: 2 Publisher: Multidisciplinary Digital Publishing Institute. [Online]. Available: \url{https://www.mdpi.com/1424-8220/22/2/550}
\BIBentrySTDinterwordspacing

\bibitem{saleh_5g-enabled_2022}
S.~Saleh, A.~S. El-Wakeel, and A.~Noureldin, ``{5G}-{Enabled} {Vehicle} {Positioning} {Using} {EKF} {With} {Dynamic} {Covariance} {Matrix} {Tuning},'' \emph{IEEE Canadian Journal of Electrical and Computer Engineering}, vol.~45, no.~3, pp. 192--198, 2022, conference Name: IEEE Canadian Journal of Electrical and Computer Engineering.

\bibitem{refsignals}
\BIBentryALTinterwordspacing
F.~Mogyorósi, P.~Revisnyei, A.~Pašić, Z.~Papp, I.~Törös, P.~Varga, and A.~Pašić, ``Positioning in 5g and 6g networks---a survey,'' \emph{Sensors}, vol.~22, no.~13, 2022. [Online]. Available: \url{https://www.mdpi.com/1424-8220/22/13/4757}
\BIBentrySTDinterwordspacing

\bibitem{bader_nlos_2022}
Q.~Bader, S.~Saleh, M.~Elhabiby, and A.~Noureldin, ``{NLoS} {Detection} for {Enhanced} {5G} {mmWave}-based {Positioning} for {Vehicular} {IoT} {Applications},'' in \emph{{IEEE} {Global} {Communications} {Conference} ({GLOBECOM})}, Dec. 2022, pp. 5643--5648.

\bibitem{shahmansoori_position_2018}
\BIBentryALTinterwordspacing
A.~Shahmansoori, G.~E. Garcia, G.~Destino, G.~Seco-Granados, and H.~Wymeersch, ``\BIBforeignlanguage{en}{Position and {Orientation} {Estimation} {Through} {Millimeter}-{Wave} {MIMO} in {5G} {Systems}},'' \emph{\BIBforeignlanguage{en}{IEEE Transactions on Wireless Communications}}, vol.~17, no.~3, pp. 1822--1835, Mar. 2018. [Online]. Available: \url{https://ieeexplore.ieee.org/document/8240645/}
\BIBentrySTDinterwordspacing

\bibitem{saleh_would_2022}
S.~Saleh, A.~Elmezayen, Q.~Bader, M.~Elhabiby, and A.~Noureldin, ``Would {Future} {mmWave} {Wireless} {Networks} {Be} an {Alternative} {Positioning} {Technique} to {GNSS}-{Based} {High} {Precision} {Positioning}?'' in \emph{Vehicular {Technology} {Conference} ({VTC})}, Jun. 2022, pp. 1--5, iSSN: 2577-2465.

\bibitem{noauthor_siradel_nodate}
\BIBentryALTinterwordspacing
``\BIBforeignlanguage{en-US}{Siradel {Simulator}}.'' [Online]. Available: \url{https://www.siradel.com/telecommunications/volcano/}
\BIBentrySTDinterwordspacing

\bibitem{NovAtel}
\BIBentryALTinterwordspacing
``Novatel - propak 6 datasheet.'' [Online]. Available: \url{https://hexagondownloads.blob.core.windows.net/public/Novatel/assets/Documents/Papers/ProPak6-PS-D18297/ProPak6-PS-D18297.pdf}
\BIBentrySTDinterwordspacing

\bibitem{TPI}
\BIBentryALTinterwordspacing
``{TPI IMU} datasheet.'' [Online]. Available: \url{https://www.mouser.ca/datasheet/2/281/s47e-522718.pdf}
\BIBentrySTDinterwordspacing

\bibitem{OBD}
\BIBentryALTinterwordspacing
``{OBDII Odometer} datasheet.'' [Online]. Available: \url{https://pdf1.alldatasheet.com/datasheet-pdf/view/542978/ELM/ELM327.html}
\BIBentrySTDinterwordspacing

\bibitem{Course}
J.~{Levine}, ``Fundamentals of {5G} small cell deployments,'' in \emph{2020 IEEE Communications Society Training Course}, 2020.

\end{thebibliography}

\begin{IEEEbiography}[{\includegraphics[width=1in,height=1.25in,clip]{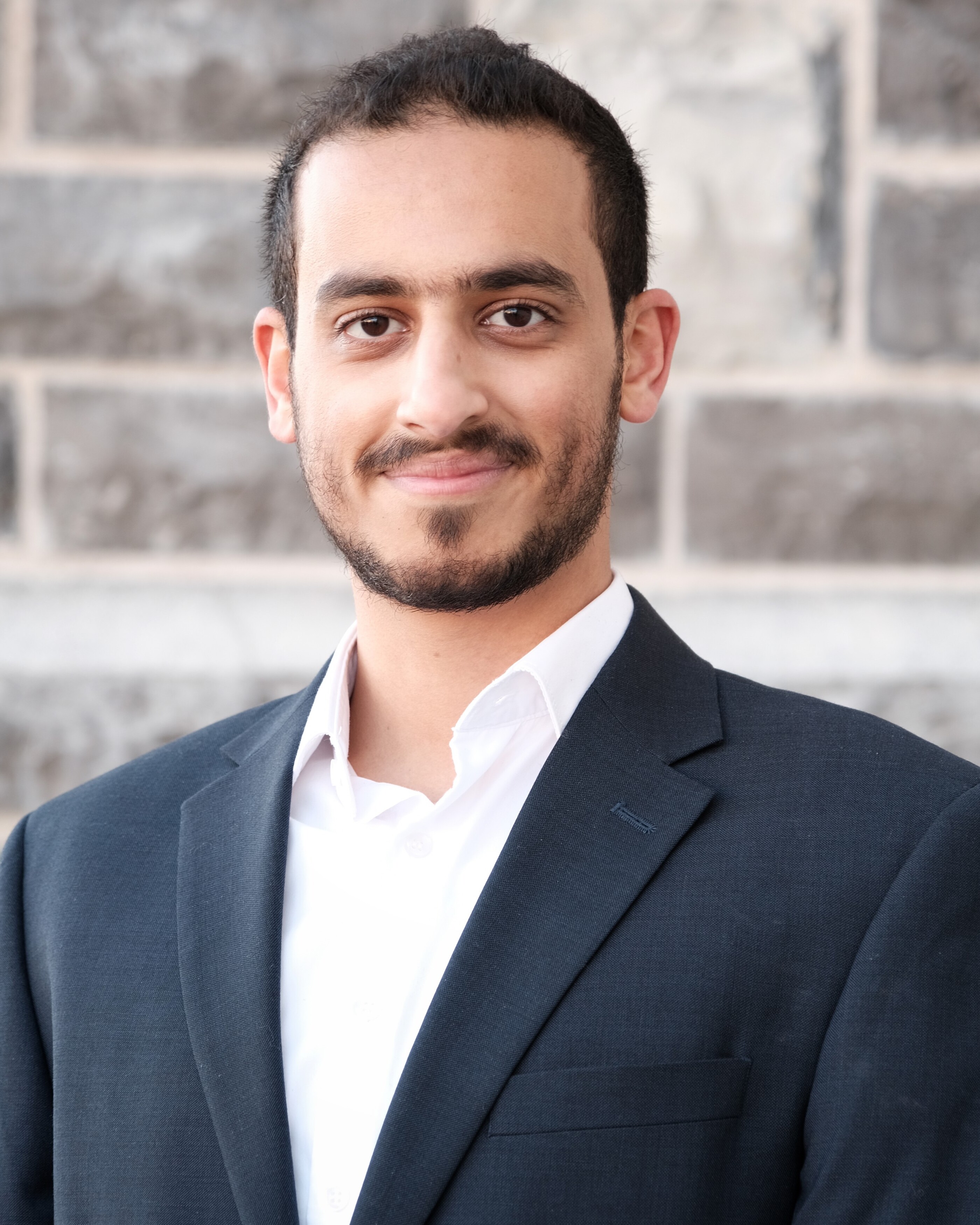}}]{Sharief Saleh} (Member, IEEE) received the B.Sc. and M.Sc. degrees in electrical engineering from Qatar University, Doha, Qatar, in 2016 and 2018, respectively, and the PhD degree in electrical engineering at Queen's University, Canada, in 2023. He is currently a post-doctoral researcher at Chalmers University of Technology. He is also a member of the Navigation and Instrumentation (NavINST) Research Lab, RMCC, Canada. His current research interests include 5G and 6G positioning, sensor fusion, estimation theory, and reinforcement learning.
\end{IEEEbiography}
\vspace{-15pt}

\begin{IEEEbiography}[{\includegraphics[width=1in,height=1.25in,clip]{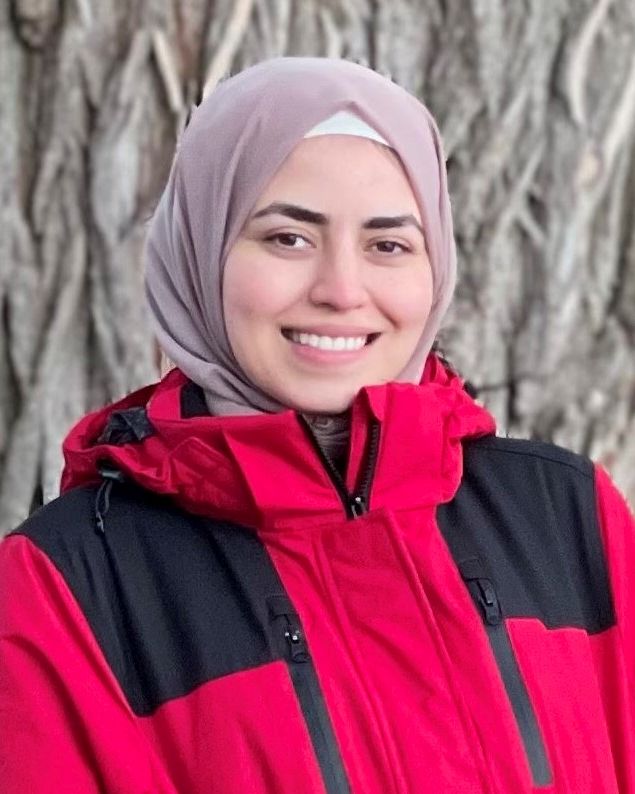}}]{Qamar Bader} (Graduate Student Member, IEEE) received her B.Sc. degree in electrical engineering from Qatar University, Doha, Qatar, in 2016 and is currently pursuing an MSc degree in electrical engineering at Queen's University, Canada. She is currently a member of the Navigation and Instrumentation (NavINST) Research Lab, RMCC. Her current research interests include 5G positioning and navigation, sensor fusion, deep learning, computer vision, and environment mapping.
\end{IEEEbiography}

\vspace{-15pt}
\begin{IEEEbiography}[{\includegraphics[width=1in ,height=1.25in ,clip]{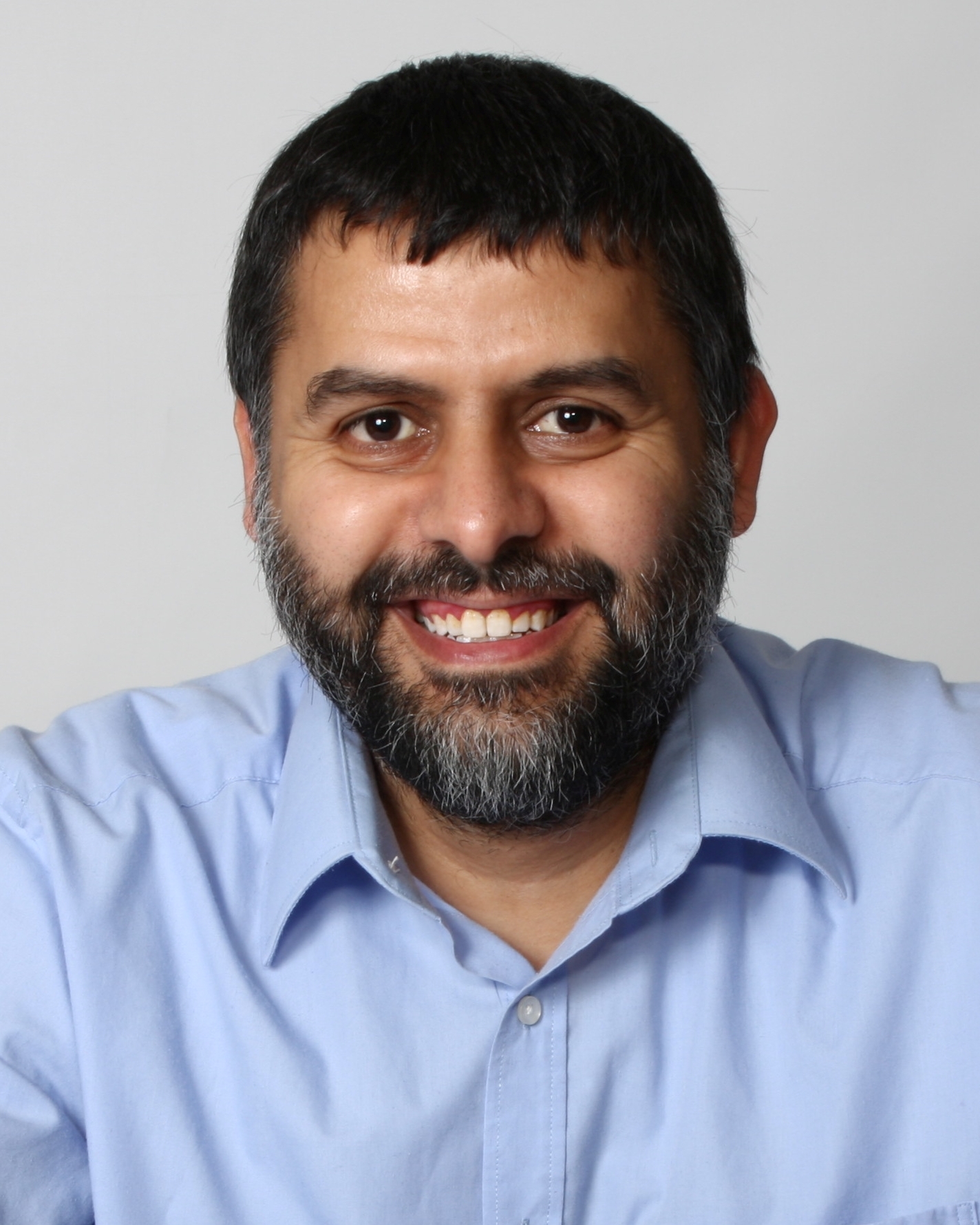}}]{Malek Karaim} (Member, IEEE) is currently a Post-Doctoral Fellow with the School of Computing, Queen’s University, Canada. He has expertise in the area of global navigation satellite systems, inertial navigation systems, and multi-sensor fusion for vehicular navigation technology. He also started recently exploring low Earth orbit (LEO) satellite-based navigation.
\end{IEEEbiography} %

\vspace{-15pt}

\begin{IEEEbiography}[{\includegraphics[width=1in,height=1.25in,clip]{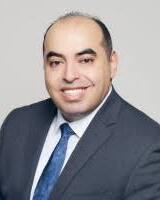}}]{Mohamed Elhabiby} was the Treasurer of the Geodesy Section at the Canadian Geophysical Union from 2008 to 2014. He is currently a Professor with the Faculty of Engineering, Ain Shams University, Cairo, Egypt. He is also the Executive Vice President and Co-Founder of Micro Engineering Tech Inc., Calgary, AB, a high-tech international company. He is a Leader of an Archaeological Mission in the Area of Great Pyramids, Cairo. He received the Astech Awards. He is named by Avenue Magazine as one of the Top 40 under 40. He is the Chair of WG 4.1.4: Imaging Techniques, Sub-Commission 4.1: Alternatives and Backups to GNSS. He chaired the Geocomputations and Cyber Infrastructure Oral Session at the Canadian Geophysical Union annual meeting 2008-2012.
\end{IEEEbiography}
\vspace{-15pt}
\begin{IEEEbiography}[{\includegraphics[width=1in,height=1.25in,clip]{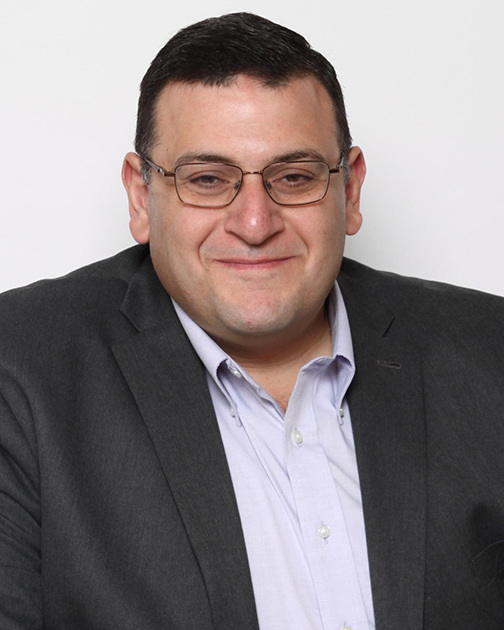}}]{Aboelmagd Noureldin} (Senior Member, IEEE) received the B.Sc. degree in electrical engineering and the M.Sc. degree in engineering physics from Cairo University, Egypt, in 1993 and 1997, respectively, and the Ph.D. degree in electrical and computer engineering from the University of Calgary, AB, Canada, in 2002. He is currently a Professor with the Department of Electrical and Computer Engineering, Royal Military College of Canada (RMCC) with a cross-appointment at the School of Computing and the Department of Electrical and Computer Engineering, Queen’s University. He is also the Founder and the Director of the Navigation and Instrumentation (NavINST) Research Lab, RMCC. He has published two books, four book chapters, and more than 270 papers in journals, magazines, and conference proceedings. His research interests include global navigation satellite systems, wireless location, and navigation, indoor positioning, and multi-sensor fusion. His research work led to 13 patents and several technologies licensed to the industry in the area of position, location, and navigation systems.
\end{IEEEbiography}
\end{document}